\documentclass[11pt]{article}
\pdfoutput=1

\usepackage{bm,wrapfig,float,array,mathtools}
\usepackage{amsfonts}
\usepackage{amssymb}
\usepackage{cite}
\usepackage{amsmath}
\usepackage{color}
\usepackage{graphicx}
\usepackage{hyperref}
\usepackage{float}
\usepackage[T1]{fontenc}
\usepackage[utf8]{inputenc}
\usepackage{alphabeta}
\usepackage{fancyvrb}
\usepackage[dvipsnames]{xcolor}
\usepackage{bold-extra}
\usepackage{lmodern}
\usepackage{cleveref}
\usepackage[normalem]{ulem}
\usepackage{tikz-cd}
\newcommand{\RNum}[1]{\uppercase\expandafter{\romannumeral #1\relax}}

\graphicspath{ {figures/} }

\usepackage{tikz}
\usetikzlibrary{arrows}
\usetikzlibrary{arrows.meta}
\usetikzlibrary{shapes.geometric,calc,arrows, positioning,shapes.misc,decorations.markings}
\tikzset{
  big arrow/.style={
    decoration={markings,mark=at position 1 with {\arrow[scale=2,#1]{>}}},
    postaction={decorate},
    shorten >=0.4pt},
  big arrow/.default=black}
  
\pgfdeclarelayer{edgelayer}
\pgfdeclarelayer{nodelayer}
\pgfsetlayers{edgelayer,nodelayer,main} 
\tikzstyle{none}=[inner sep=0pt] 

\tikzstyle{NodeCross}=[draw, shape=circle, cross out, inner sep=0pt, minimum size=6pt,line width=0.25mm]
\tikzstyle{Circle}=[draw, shape=circle, black, inner sep=0pt, minimum size=6pt]
\tikzstyle{rtriangle}=[fill=black, regular polygon, regular polygon sides=3, rotate=90, inner sep=0pt, minimum size=8pt]
\tikzstyle{ltriangle}=[fill=black, regular polygon, regular polygon sides=3, rotate=270, inner sep=0pt, minimum size=8pt]
\tikzstyle{rtriangleblue}=[fill={rgb,255: red,17; green,160; blue,255}, regular polygon, regular polygon sides=3, rotate=90, inner sep=0pt, minimum size=8pt]
\tikzstyle{ltriangleblue}=[fill={rgb,255: red,17; green,160; blue,255}, regular polygon, regular polygon sides=3, rotate=270, inner sep=0pt, minimum size=8pt]
\tikzstyle{rtrianglegreen}=[fill={rgb,255: red,69; green,255; blue,28}, regular polygon, regular polygon sides=3, rotate=90, inner sep=0pt, minimum size=8pt]
\tikzstyle{ltrianglegreen}=[fill={rgb,255: red,69; green,255; blue,28}, regular polygon, regular polygon sides=3, rotate=270, inner sep=0pt, minimum size=8pt]
\tikzstyle{Uprtriangle}=[fill=black, regular polygon, regular polygon sides=3, rotate=0, inner sep=0pt, minimum size=8pt]
\tikzstyle{Downltriangle}=[fill=black, regular polygon, regular polygon sides=3, rotate=180, inner sep=0pt, minimum size=8pt]
\tikzstyle{rtriangleAmber}=[fill={rgb,255: red, 191; green, 144; blue, 63}, regular polygon, regular polygon sides=3, rotate=90, inner sep=0pt, minimum size=8pt]
\tikzstyle{UprtriangleViolett}=[fill={rgb,255: red,255; green,0; blue,0}, regular polygon, regular polygon sides=3, rotate=0, inner sep=0pt, minimum size=8pt]
\tikzstyle{ltrianglered}=[fill={rgb,255: red,191; green,0; blue,0}, regular polygon, regular polygon sides=3, rotate=270, inner sep=0pt, minimum size=8pt]

\tikzstyle{Downltriangle}=[fill=black, regular polygon, regular polygon sides=3, rotate=180, inner sep=0pt, minimum size=8pt]
\tikzstyle{UpRighttriangle}=[fill=black, regular polygon, regular polygon sides=3, rotate=45, inner sep=0pt, minimum size=8pt]
\tikzstyle{UpLefttriangle}=[fill=black, regular polygon, regular polygon sides=3, rotate=315, inner sep=0pt, minimum size=8pt]
\tikzstyle{DownRighttriangle}=[fill=black, regular polygon, regular polygon sides=3, rotate=135, inner sep=0pt, minimum size=8pt]
\tikzstyle{DownLighttriangle}=[fill=black, regular polygon, regular polygon sides=3, rotate=225, inner sep=0pt, minimum size=8pt]

\tikzstyle{Star}=[draw, shape=star, fill=black, star points=8, inner sep=0pt, minimum size=8pt]

\tikzstyle{DashedLine}=[-, densely dashed, line width=0.25mm]
\tikzstyle{DashedLineBrown}=[-, densely dashed, line width=0.25mm, draw={rgb,255: red,155; green,103; blue,51}]
\tikzstyle{DashedLineFall}=[-, densely dashed, line width=0.25mm, draw={rgb,255: red,195; green,0; blue,0}]
\tikzstyle{DashedLineViolett}=[-, densely dashed, line width=0.25mm, draw={rgb,255: red,139; green,41; blue,148}]
\tikzstyle{DottedLine}=[-, dotted, line width=0.25mm]
\tikzstyle{BlueLine}=[-, fill=none, draw={rgb,255: red,17; green,160; blue,255}, line width=0.25mm]
\tikzstyle{GreenLine}=[-, fill=none, draw={rgb,255: red,69; green,255; blue,28}, line width=0.25mm]
\tikzstyle{RedLine}=[-, draw={rgb,255: red,191; green,0; blue,0}, fill=none, line width=0.25mm]
\tikzstyle{LBRedLine}=[-, draw={rgb,255: red,255; green,0; blue,0}, fill=none, line width=0.25mm]
\tikzstyle{DashedLineRed}=[-, densely dashed, fill=none, draw={rgb,255: red,191; green,0; blue,0}, line width=0.25mm]
\tikzstyle{ThickLine}=[-, line width=0.25mm]
\tikzstyle{ViolettLine}=[-, draw={rgb,255: red,132; green,60; blue,191}, fill=none, line width=0.25mm]
\tikzstyle{ViolettDashedLine}=[-, densely dashed, draw={rgb,255: red,132; green,60; blue,191}, fill=none, line width=0.25mm]
\tikzstyle{AmberLine}=[-, draw={rgb,255: red,191; green,144; blue,63}, fill=none, line width=0.25mm]
\tikzstyle{DashedRedThick}=[-, densely dashed, fill=none, draw={rgb,255: red,191; green,0; blue,0}, line width=0.40mm]
\tikzstyle{DashedBlueThick}=[-, densely dashed, fill=none, black, line width=0.40mm]
\tikzstyle{DottedLineRed}=[-, dotted, line width=0.25mm, draw={rgb,255: red,191; green,0; blue,0}]
\tikzstyle{DottedLineBlue}=[-, dotted, line width=0.25mm, draw={rgb,255: red,17; green,160; blue,255}]
\tikzstyle{ArrowLineRight}=[-, -{Stealth[scale=1.75]}, line width=0.1mm, scale=5]
\tikzstyle{ArrowLineLeft}=[-, {Stealth[scale=1.75]}-, line width=0.1mm, scale=5]
\tikzstyle{ArrowLineRightBlue}=[-, -{Stealth[scale=1.75]}, line width=0.1mm, draw={rgb,255: red,17; green,160; blue,255}]
\tikzstyle{ArrowLineLeftBlue}=[-, {Stealth[scale=1.75]}-, line width=0.1mm, draw={rgb,255: red,17; green,160; blue,255}]
\tikzstyle{ArrowLineRightRed}=[-, -{Stealth[scale=1.75]}, line width=0.1mm, draw={rgb,255: red,191; green,0; blue,0}]
\tikzstyle{ArrowLineLeftRed}=[-, {Stealth[scale=1.75]}-, line width=0.1mm, draw={rgb,255: red,191; green,0; blue,0}]

\newcommand{\bea}{\begin{eqnarray}}
\newcommand{\eea}{\end{eqnarray}}
\newcommand{\be}{\begin{equation}}
\newcommand{\ee}{\end{equation}}
\newcommand{\ba}{\begin{aligned}}
\newcommand{\ea}{\end{aligned}}
\newcommand{\bit}{\begin{itemize}}
\newcommand{\eit}{\end{itemize}}
\newcommand{\ben}{\begin{enumerate}}
\newcommand{\een}{\end{enumerate}}

\newcommand{\wh}{\widehat}

\newcommand{\half}{\frac{1}{2}}

\newcommand{\Z}{{\mathbb Z}}

\renewcommand{\P}{{\mathbb P}}

\newcommand{\cF}{\mathcal{F}}
\newcommand{\cG}{\mathcal{G}}

\newcommand{\cN}{\mathcal{N}}

\newcommand{\cP}{\mathcal{P}}

\newcommand{\F}{\mathsf{F}}
\renewcommand{\S}{\mathsf{S}}

\newcommand{\A}{\mathsf{A}}

\newcommand{\C}{\mathsf{C}}

\renewcommand{\L}{\mathsf{\Lambda}}

\newcommand{\fT}{\mathfrak{T}}

\newcommand{\fe}{\mathfrak{e}}
\newcommand{\ff}{\mathfrak{f}}
\newcommand{\fg}{\mathfrak{g}}
\newcommand{\fh}{\mathfrak{h}}
\newcommand{\su}{\mathfrak{su}}
\renewcommand{\sp}{\mathfrak{sp}}
\newcommand{\so}{\mathfrak{so}}
\renewcommand{\u}{\mathfrak{u}}

\newcommand{\bF}{{\mathbb F}}

\begin{document}

\baselineskip=18pt  
\numberwithin{equation}{section}  
\allowdisplaybreaks  

\thispagestyle{empty}

\vspace*{0.8cm} 
\begin{center}
{{\huge Discovering T-Dualities of Little String Theories}}

\vspace*{1.5cm}
Lakshya Bhardwaj\\

\vspace*{.5cm} 
{\it  Mathematical Institute, University of Oxford, \\
Andrew-Wiles Building,  Woodstock Road, Oxford, OX2 6GG, UK}

\vspace*{0.8cm}
\end{center}
\vspace*{.5cm}

\noindent
We describe a general method for deducing T-dualities of little string theories, which are dualities between these theories that arise when they are compactified on circle. The method works for both untwisted and twisted circle compactifications of little string theories and is based on surface geometries associated to these circle compactifications. The surface geometries describe information about Calabi-Yau threefolds on which M-theory can be compactified to construct the corresponding circle compactified little string theories. Using this method, we deduce at least one T-dual, and in some cases multiple T-duals, for untwisted and twisted circle compactifications of most of the little string theories that can be described on their tensor branches in terms of a 6d supersymmetric gauge theory with a simple non-abelian gauge group, which are also known as rank-0 little string theories. This includes little string theories carrying $\cN=(1,1)$ and $\cN=(1,0)$ supersymmetries. For many, but not all, circle compactifications of $\cN=(1,1)$ little string theories, we have T-dualities that realize Langlands dualities between affine Lie algebras. Along the way, we find another discrete theta angle distinct from 0 and $\pi$ for an E-string node.

\newpage

\tableofcontents

\section{Introduction}
Little string theories (LSTs) are fascinating quantum-mechanical theories. They are expected to be UV complete, but are neither local quantum field theories (QFTs) nor fully-fledged string theories, sitting somewhere in between the two. Unlike string theories, gravity is not dynamical in these theories, and as a consequence they behave very similarly to local QFTs. The key property that differentiates LSTs from local QFTs is that LSTs admit T-dualities \cite{Seiberg:1997zk,Intriligator:1997dh,Intriligator:1999cn},
which is the topic of this work.

All the known LSTs exist in six spacetime dimensions, and carry at least $\cN=(1,0)$ supersymmetry. Some very special LSTs carry $\cN=(2,0)$ or $\cN=(1,1)$ supersymmetries. A uniform definition for all the known LSTs is provided as compactifications of F-theory on (a special class of) elliptically fibered Calabi-Yau threefolds. This includes both the standard F-theory compactifications \cite{Bhardwaj:2015oru}, and also the recently explored \textit{frozen} F-theory compactifications \cite{Bhardwaj:2019hhd,Bhardwaj:2018jgp}. These theories can also be classified using anomaly cancellation \cite{Seiberg:1996qx,Danielsson:1997kt,Bhardwaj:2015xxa}.

To study T-dualities\footnote{See \cite{Hohenegger:2015btj,Bhardwaj:2015oru,Hohenegger:2016eqy,Hohenegger:2016yuv,Bastian:2017ing,Bastian:2017ary,Bastian:2018dfu,Bastian:2018jlf} for recent works on little string T-dualities.} of such LSTs, we need to compactify the F-theory setups on circle, which are dual to M-theory compactified on the same elliptically fibered Calabi-Yau threefolds\footnote{We also study twisted circle compactifications of LSTs, which involve turning on non-zero discrete holonomies along the compactification circle. In such cases, the M-theory threefold differs from the F-theory threefold.}. T-dualities thus express themselves as geometric isomorphisms between the corresponding threefolds. 

In this paper, we study such geometric isomorphisms in terms of graphical structures that we call \textit{\textbf{surface geometries}}, which capture the information associated to holomorphic curves and surfaces inside the threefold. Surface geometries have been immensely useful in the recent studies on 5d and 6d superconformal field theories (SCFTs) \cite{Jefferson:2018irk,Bhardwaj:2018yhy,Bhardwaj:2018vuu,Bhardwaj:2019ngx,Bhardwaj:2019jtr,Bhardwaj:2019fzv,Bhardwaj:2019xeg,Bhardwaj:2020gyu,Bhardwaj:2020kim,Bhardwaj:2020phs,Bhardwaj:2020ruf,Bhardwaj:2020avz,Apruzzi:2021vcu,Kim:2020hhh,Hayashi:2021pcj,Kim:2021gyj,Hayashi:2018lyv,Apruzzi:2018nre,Apruzzi:2019opn,Apruzzi:2019kgb,Saxena:2020ltf,Eckhard:2020jyr,Morrison:2020ool,Hubner:2020uvb,Albertini:2020mdx,Closset:2020scj,Braun:2021lzt,Tian:2021cif,Closset:2021lwy}. They were proposed as useful tools for these studies in \cite{Jefferson:2018irk}, while the essential graphical calculus for these purposes was developed in \cite{Bhardwaj:2018yhy,Bhardwaj:2018vuu,Bhardwaj:2019ngx,Bhardwaj:2019jtr,Bhardwaj:2019fzv,Bhardwaj:2020kim,Bhardwaj:2020ruf,Bhardwaj:2020avz}. The present work is thus part of the vast progress \cite{Jefferson:2017ahm,Jefferson:2018irk,Ashok:2017bld,Hayashi:2018lyv,Cheng:2018wll,Bhardwaj:2018yhy,Cabrera:2018jxt,Bastian:2018fba,Chaney:2018gjc,Bhardwaj:2018vuu,Apruzzi:2018nre,Closset:2018bjz,Hayashi:2019yxj,Bhardwaj:2019hhd,Apruzzi:2019vpe,Apruzzi:2019opn,Kim:2019dqn,Kim:2019uqw,Uhlemann:2019ypp,Bhardwaj:2019ngx,Apruzzi:2019enx,Bhardwaj:2019jtr,Bhardwaj:2019fzv,Cota:2019cjx,Bhardwaj:2019xeg,Apruzzi:2019kgb,Gu:2019pqj,Uhlemann:2019lge,Hayashi:2019jvx,Closset:2019juk,Saxena:2020ltf,Bhardwaj:2020gyu,Bourget:2020gzi,Eckhard:2020jyr,Crichigno:2020ouj,Bhardwaj:2020kim,Morrison:2020ool,Uhlemann:2020bek,Hubner:2020uvb,Gu:2020fem,Albertini:2020mdx,Collinucci:2020jqd,Closset:2020scj,Akhond:2020vhc,vanBeest:2020kou,Apruzzi:2020zot,Bhardwaj:2020phs,Bhardwaj:2020ruf,Bhardwaj:2020avz,Fluder:2020pym,Bergman:2020myx,VanBeest:2020kxw,Closset:2020afy,Hayashi:2020hhb,Kim:2020hhh,Kim:2021cua,Hayashi:2021pcj,Duan:2021ges,Braun:2021lzt,Kim:2021gyj,Cvetic:2021sxm,vanBeest:2021xyt,Acharya:2021jsp,Apruzzi:2021mlh,Tian:2021cif,Closset:2021lwy,Saidi:2021kan,Kim:2021fxx,DelZotto:2022fnw,Kim:2022dbr,Akhond:2022jts,Cvetic:2022imb,DeMarco:2022dgh,Jia:2022dra,Apruzzi:2021vcu} made recently in understanding 5d and 6d supersymmetric UV complete theories -- which so far has included SCFTs and KK theories; the present paper adds LSTs to this list.

The strategy for using surface geometries to determine T-dualities of an LST $\fT$ can be summarized as follows:
\ben
\item Use the tensor branch description of $\fT$. This can be packaged in terms of a graph with nodes and edges. This graph is used in the following operations for studying T-duals of the untwisted compactification of $\fT$. For studying T-duals of a twisted compactification of $\fT$, one uses information regarding the twist to convert the graph for $\fT$ into a new graph, which is then used in the following operations. Let us denote the circle compactified theory being studied as $\fT_{S^1}^\rho$, where $\rho$ denotes the twist involved, which takes trivial value if the circle compactification being studied is untwisted.
\item To each node in the graph associated to $\fT_{S^1}^\rho$, we can associate a collection of Hirzebruch surfaces (i.e. compact Kahler surfaces with a choice of $\P^1$ fibration) glued to each other.
\item If two nodes are connected by an edge, we glue together the collections of Hirzebruch surfaces associated to the two nodes. By this point, we have described a surface geometry associated to $\fT_{S^1}^\rho$.
\item Some Hirzebruch surfaces contain multiple $\P^1$ fibers. Changing the choice of $\P^1$ fibers judiciously lets us recognize the resulting surface geometry as a surface geometry associated to $\wh\fT^{\wh\rho}_{S^1}$, i.e. a $\wh\rho$ (which can take a trivial value for untwisted compactification) twisted circle compactification of an LST $\wh\fT$. This means that we have a T-duality between LSTs $\fT$ and $\wh\fT$, where $\fT$ is compactified with twist $\rho$ and $\wh\fT$ is compactified with twist $\wh\rho$.
\een

We illustrate this procedure by finding a T-dual for twisted and untwisted compactifications of many rank-0 LSTs describable on their tensor branches as 6d $\cN=(1,0)$ gauge theories with simple gauge algebras. More concretely, we study T-duals of the following systems:
\bit
\item 6d theory with any simple gauge algebra $\fg$ and a hyper in adjoint representation $\A$. This theory has $\cN=(1,1)$ supersymmetry. For $\fg=\su(n)$, we can turn on a rational theta angle $\theta=2\pi \frac pq$ \cite{Witten:1997kz}, and for $\fg=\sp(n)$, we can turn on a $\Z_2$ valued theta angle $\theta=0,\pi$. We provide a T-dual for each untwisted compactification of these theories for $\fg\neq\su(n)$. On the other hand, for $\fg=\su(n)$, we provide a T-dual for untwisted compactification only of $\theta=0$ theory\footnote{We leave the discussion of T-dualities for non-zero theta angles to future works. It is an open question whether there exist surface geometries for these LSTs, or if some modifications are needed.}.\\
For $\fg=\su(n),\so(2n),\fe_6$ we have the possibility of twisting the compactification by the $\Z_2$ outer-automorphism of such a Lie algebra $\fg$. For $\fg=\su(n)$, we must have $\theta=0,\pi$ for the twist to be allowed. We provide a T-dual for each such twisted compactification.\\
For $\fg=\so(8)$, we have an additional possibility of twisting the compactification by the $\Z_3$ outer-automorphism of $\so(8)$, and we provide a T-dual for this twisted compactification.
\item 6d theory with $\su(n\ge3)$ gauge algebra, 2 hypers in antisymmetric irrep $\L^2$ and 16 hypers in fundamental irrep $\F$. We have the possibility of twisting the compactification by the $\Z_2$ outer-automorphism of $\su(n)$. We provide a T-dual for each such compactification.
\item 6d theory with $\su(2)$ gauge algebra and 16 hypers in fundamental irrep $\F$. We provide a T-dual for the untwisted compactification.
\item 6d theory with $\su(n\ge3)$ gauge algebra, 1 hyper in antisymmetric irrep $\L^2$ and 1 hyper in symmetric irrep $S^2$. We have the possibility of twisting the compactification by the $\Z_2$ outer-automorphism of $\su(n)$. We provide a T-dual only for the untwisted compactification.
\item 6d theory with $\su(6)$ gauge algebra, 1 hyper in three-index antisymmetric irrep $\L^3$ and 18 hypers in fundamental irrep $\F$. We have the possibility of twisting the compactification by the $\Z_2$ outer-automorphism of $\su(6)$. We provide a T-dual for each such compactification.

Note that in the T-dual for the untwisted compactification, we find a hitherto unknown discrete theta angle for an E-string $\sp(0)$ node, that is distinct from $\theta=0,\pi$. We call this as a \textbf{new theta angle}. See equation (\ref{new}). The distinction between the various theta angles is field-theoretically described in terms of different embeddings of the gauge symmetries into the full $\fe_8$ flavor symmetry of the E-string theory, leaving behind different residual flavor symmetries. In this case, while the $\theta=0,\pi$ have residual flavor symmetries $\su(2)$ and $\u(1)$ respectively, the new theta angle has residual flavor symmetry $\su(3)$. The residual flavor symmetries can be determined using the respective surface geometries by finding $(-2)$-curves inside the compact surfaces that have zero intersection with all the gluing curves, which means that they can be attached to $\P^1$ fibers of non-compact Hirzebruch surfaces describing (non-abelian) flavor symmetries \cite{Bhardwaj:2020ruf,Bhardwaj:2020avz}. The intersection pattern of these $(-2)$-curves determines the intersection pattern of the non-compact Hirzebruch surfaces, thus determining the form of the flavor symmetry algebra.
\item 6d theory with $\su(6)$ gauge algebra, 1 half-hyper in three-index antisymmetric irrep $\L^3$, 1 hyper in antisymmetric irrep $\L^2$ and 17 hypers in fundamental irrep $\F$. We have the possibility of twisting the compactification by the $\Z_2$ outer-automorphism of $\su(6)$. We provide a T-dual for each such compactification.
\item 6d theory with $\so(2n+1)$ gauge algebra for $3\le n\le 6$, $2^{6-n}$ hypers in spinor irrep $\S$ and $2n-3$ hypers in vector irrep $\F$. 
We provide a T-dual for the untwisted compactifications.
\item 6d theory with $\so(8)$ gauge algebra, 4 hypers in spinor irrep $\S$, 4 hypers in cospinor irrep $\C$ and 4 hypers in vector irrep $\F$. We have the possibility of twisting the compactification by either the $\Z_2$ outer-automorphism of $\so(8)$ or by the $\Z_3$ outer-automorphism of $\so(8)$. We provide a T-dual for each such compactification.
\item 6d theory with $\so(10)$ gauge algebra, 4 hypers in spinor irrep $\S$ and 6 hypers in vector irrep $\F$. We have the possibility of twisting the compactification by the $\Z_2$ outer-automorphism of $\so(10)$. We provide a T-dual for each such compactification.
\item 6d theory with $\so(12)$ gauge algebra, 2 hypers in spinor irrep $\S$ and 8 hypers in vector irrep $\F$. 
We provide a T-dual for the untwisted compactification.
\item 6d theory with $\so(12)$ gauge algebra, 1 hyper in spinor irrep $\S$, 1 hyper in cospinor irrep $\C$ and 8 hypers in vector irrep $\F$. We have the possibility of twisting the compactification by the $\Z_2$ outer-automorphism of $\so(12)$. 
We provide a T-dual for each such compactification.
\item 6d theory with $\so(14)$ gauge algebra, 1 hyper in spinor irrep $\S$ and 10 hypers in vector irrep $\F$. We have the possibility of twisting the compactification by the $\Z_2$ outer-automorphism of $\so(14)$. We provide a T-dual for each such compactification.
\item 6d theory with $\sp(n\ge2)$ gauge algebra, 1 hyper in antisymmetric irrep $\L^2$ and 16 hypers in fundamental irrep $\F$. We provide a T-dual for the untwisted compactification.
\item 6d theory with $\sp(3)$ gauge algebra, 1 half-hyper in three-index antisymmetric irrep $\L^3$ and 35 half-hypers in fundamental irrep $\F$. We provide a T-dual for the untwisted compactification.
\item 6d theory with $\fg_2$ gauge algebra and 10 hypers in 7-dimensional irrep $\F$. We provide a T-dual for the untwisted compactification.
\eit

\section{Surface Geometries for Circle Compactifications of LSTs}
In this paper, we will use the conventions and notations for surfaces that were introduced in \cite{Bhardwaj:2019fzv}.

\subsection{Graphs Associated to Circle Compactifications of LSTs}
Consider, as in the introduction, a circle compactification $\fT_{S^1}^\rho$ of a LST $\fT$ twisted by $\rho$. To such a circle compactification, one can associate a graph which essentially encodes the information regarding the twist and the low-energy theory arising on the tensor branch of the LST. The construction of the graph was explained in \cite{Bhardwaj:2019fzv} for circle compactifications of 6d SCFTs, and carries over in a similar fashion to circle compactifications of LSTs.

We will denote the nodes of the graph by $i$. Each node carries a non-negative integer $k_i$ and a simple affine Lie algebra $\fg_i^{(q_i)}$ where $\fg_i$ is a simple finite Lie algebra. The affine algebra can be trivial, and there are two possible choices denoted by $\sp(0)^{(1)}$ and $\su(1)^{(1)}$. See \cite{Bhardwaj:2019fzv} for more details.

An edge between two nodes can be directional or non-directional, and carries a positive integer label. Using the information about $k_i$ and the edges, we can extract a matrix $[\Omega^{ij}]$ as described in \cite{Bhardwaj:2019fzv}. For a circle compactification of a 6d SCFT $[\Omega^{ij}]$ is a positive-definite matrix, while for a circle compactification of an LST, $[\Omega^{ij}]$ is a positive semi-definite matrix with a 1-dimensional space of null vectors.

From the structure of the graph (including some additional decorations of nodes and edges), we can also extract a representation $R_\fT^\rho$ of a finite Lie algebra
\be
\fh=\bigoplus_i\fh_i
\ee
where $\fh_i$ is the subalgebra (which is tabulated in \cite{Bhardwaj:2019fzv}) of $\fg_i$ left invariant by the order $q_i$ outer-automorphism used in the twist. The physical interpretation of $R_\fT^\rho$ is that the circle compactification $\fT^\rho_{S^1}$ of the 6d LST $\fT$ admits a Coulomb branch phase in which at extremely low energies we find a 5d gauge theory with $\fh$ gauge algebra and hypers transforming in representation $R_\fT^\rho$ of $\fh$.

\subsection{Surface Geometries from Graphs}
The information about the graph associated to $\fT^\rho_{S^1}$ can be converted into surface geometries describing various 5d $\cN=1$ supersymmetric Coulomb branch phases of $\fT^\rho_{S^1}$.
Such a phase $\cP$ of $\fT^\rho_{S^1}$ can be constructed by compactifying M-theory on a 
smooth Calabi-Yau threefold $X_\cP$. 
A 5d $\cN=1$ abelian gauge theory $\fT_\cP$ with massive matter content describes the low-energy theory at a point in the interior of the phase $\cP$. This gauge theory $\fT_\cP$ can be identified as the low-energy theory obtained upon compactification of M-theory on $X_\cP$.

Each $X_\cP$ corresponds to a different surface geometry $\cG_\cP$. Surface geometries $\cG_\cP$ and $\cG_{\cP'}$ for two different phases $\cP$ and $\cP'$ are related by a sequence of flop transitions. In this work, we will study phases for which the corresponding threefold $X_\cP$ is a genus-one fibration. We will call such phases as \textit{genus-one phases}. There always exists at least one genus-one phase for each $\fT^\rho_{S^1}$.

The surface geometry for any genus-one phase takes the following form. For each $i$, we have a collection of surfaces $S_{i,\alpha}$ where $0\le\alpha\le r_i$ and $r_i$ is the rank of $\fh_i$. Consider first the case when $\fh_i$ is non-trivial, i.e. when $r_i>0$. In this case, each $S_{i,\alpha}$ is a Hirzebruch surface with (typically non-generic) blowups. Let $f_{i,\alpha}$ be the $\P^1$ fiber of the Hirzebruch surface\footnote{For us, a Hirzebruch surface is a surface with a chosen $\P^1$ fibration, which we denote by $f$. There might be other $\P^1$ fibers in the surface, which give different Hirzebruch surface descriptions for the same surface.} $S_{i,\alpha}$. These Hirzebruch surfaces must intersect such that we have
\be
-f_{i,\alpha}\cdot S_{j,\beta}=\delta_{ij}C_{i,\alpha\beta}
\ee
where $\delta_{ij}$ is the Kronecker delta and $C_{i,\alpha\beta}$ is the Cartan matrix for $\fg_i^{(q_i)}$. Thus each $S_{i,\alpha}$ corresponds to a node in the Dynkin diagram of $\fg_i^{(q_i)}$, and our convention is that $S_{i,0}$ corresponds to the affine node, i.e. the nodes $S_{i,\alpha}$ for $\alpha\neq0$ describe the Dynkin diagram of $\fh_i$.

The curve
\be
f_i:=\sum_\alpha d_{i,\alpha}f_{i,\alpha}
\ee
where $d_{i,\alpha}$ is the row null vector of the Cartan matrix of $\fg^{(q_i)}_i$ with minimal positive integer values, is a degenerate genus-one fiber.

For trivial $\fh_i$, we have a single surface $S_{i,0}$ which arises only for the following cases:
\bit
\item $\Omega^{ii}=1$ without loop, in which case the surface $S_{i,0}$ is a Hirzebruch surface with 8 blowups, which admits an elliptic fibration. There are various ways of representing such a surface. The common representations used in this paper are as follows:
\bit
\item If we represent it as $\bF_0^8$, then the elliptic fiber is $f_i=2e+2f-\sum x_i$.
\item If we represent it as $\bF_1^8$, then the elliptic fiber is $f_i=2h+f-\sum x_i$.
\item Finally, if we represent it as $\bF_2^8$, then the elliptic fiber is $f_i=2h-\sum x_i$.
\eit
In these cases, we define a curve $E_{i,0}=f-x_1$ to be used later. The (trivial) algebra attached to such a node is $\fg_i^{(q_i)}=\sp(0)^{(1)}$.
\item $\Omega^{ii}=2$ without loop, in which case the surface $S_{i,0}$ is a self-glued Hirzebruch surface which admits an elliptic fibration. In this paper, we typically use the following ways of representing this surface:
\bit
\item $\bF_0^2$ with the two blowups $x$ and $y$ glued to each other. The elliptic fiber is $f_i=e+f-x-y$.
\item $\bF_0^2$ with $e-x$ and $e-y$ glued to each other. The elliptic fiber is $f_i=f$.
\eit
In both of these cases, we define a curve $E_{i,0}=e$ to be used later. The (trivial) algebra attached to such a node is $\fg_i^{(q_i)}=\su(1)^{(1)}$.
\item $\Omega^{ii}=2$ with a loop, in which case the surface $S_{i,0}$ is a self-glued Hirzebruch surface which in this paper, we represent this surface as $\bF_1^2$ with the two blowups $x$ and $y$ glued to each other. It admits a special genus-one fiber $f_i=2h+f-2x-2y$. We define a curve $E_{i,0}=e$ to be used later. The (trivial) algebra attached to such a node is $\fg_i^{(q_i)}=\su(1)^{(1)}$.
\item It is also possible to have $\Omega^{ii}=1$ with a loop, which we do not discuss in this paper. The (trivial) algebra attached to such a node is $\fg_i^{(q_i)}=\sp(0)^{(1)}$.
\eit

The curve $f_i$ satisfies the property that
\be
f_i\cdot S_{j,\beta}=0
\ee
for all $i,j,\beta$.

We have the property
\be
f_{i,\alpha}\cdot S_j=0
\ee
for all $i,\alpha,j$ where for $\fh_j$ non-trivial
\be
S_j:=\sum_{\beta}d^\vee_{j,\beta}S_{j,\beta}
\ee
and $d^\vee_{j,\beta}$ is a null column vector of the Cartan matrix of $\fg_j^{(q_j)}$ with minimal positive integer values, $S_i=S_{i,0}$ for trivial $\fh_i$, and $f_{i,0}:=f_i$ (i.e. the elliptic fiber, not the $\P^1$ fiber of the Hirzebruch surface) for trivial $\fh_i$.

Additionally we have the property
\be
-S_i\cdot E_{j,0}=\Omega^{ij}
\ee
where
where $E_{j,0}=e_{j,0}$ is the base curve of the Hirzebruch surface $S_{j,0}$ if $\fh_j$ is non-trivial, and $E_{j,0}$ for $\fh_j$ trivial was defined above.

The blowups living in the surfaces $S_{i,\alpha}$ for non-trivial $\fh_i$ describe hypermultiplets charged under the gauge algebra $\fh=\oplus_i\fh_i$. The map between the blowups and representations of hypers is as follows. Pick a blowup $x$ of a surface $S_{i,\alpha}$ such that $\fh_i$ is non-trivial. It lives in a family $\cF$ of blowups, where some other blowup $y$ of some other surface $S_{j,\beta}$ (such that $\fh_j$ is non-trivial) is in the same family $\cF$ if and only if one of the following conditions hold:
\bit
\item $x$ is glued to $y$.
\item $x$ is glued to $f-y$.
\item $f-x$ is glued to $y$.
\item If $(i,\alpha)=(j,\beta)$, i.e. if $x$ and $y$ live in the same surface, and $f_{i,\alpha}-x-y$ is glued to $f_{k,\gamma}$ for some other surface $S_{k,\gamma}$.
\item If $(i,\alpha)=(j,\beta)$, i.e. if $x$ and $y$ live in the same surface, and $x-y$ is glued to $f_{k,\gamma}$ for some other surface $S_{k,\gamma}$.
\eit
Such a family $\cF$ of blowups describes a hypermultiplet living in a representation $R_\cF$ of $\fh$. The information about this representation can be described in terms of its weights which are obtained as follows. For every blowup $x\in\cF$ compute its intersections 
\be
-x\cdot S_{k,\gamma}
\ee
with all surfaces $S_{k,\gamma}$ with $\gamma\neq0$ for non-trivial $\fh_k$. These intersection numbers describe the Dynkin coefficients for a weight $w_x$ of $R_\cF$. Collect all weights $w_x$ for all blowups $x\in\cF$. Then $R_\cF$ is recovered as the representation with minimal number of weights that contain the weights $w_x$ for all $x\in\cF$. The full hypermultiplet content carried by the 5d gauge theory associated to $\fT^\rho_{S^1}$ is recovered by combining the representations $R_\cF$ for all families $\cF$.

The intersection number
\be
x\cdot S_{k,0}
\ee
for a blowup $x$ living in surface $S_{i,\alpha}$ with non-trivial $\fh_i$, and for all $k$, is constrained by the property that
\be
x\cdot S_{k}=0
\ee

Finally, the gluings of curves are constrained such that taking combinations of the gluing rules we find that a representative of a multiple of $f_i$ is glued to a representative of a multiple of $f_j$ if there is an edge between nodes $i$ and $j$. More precisely,
\bit
\item If there is an undirected edge carrying a label $n$ between $i$ and $j$, then the gluing rules imply the gluing
\be
n f_i\sim n f_j
\ee
but not $mf_i\sim mf_j$ for any $m<n$.
\item If there is a directed edge from $i$ to $j$ carrying label $n$, then the gluing rules imply the gluing
\be
f_i\sim n f_j
\ee
\eit

\subsection{T-dualities from local S-dualities}
Consider a circle compactification $\fT_{1,S^1}^{\rho_1}$ of an LST $\fT_1$ and a circle compactification $\fT_{2,S^1}^{\rho_2}$ of an LST $\fT_2$. We say that  $\fT_{1,S^1}^{\rho_1}$ and $\fT_{2,S^1}^{\rho_2}$ are \textit{T-dual} if we find a surface geometry $\cG$ that describes both $\fT_{1,S^1}^{\rho_1}$ and $\fT_{2,S^1}^{\rho_2}$.

In this paper, we will only consider a subset of such T-dualities which can be described as follows. Let $\cG_1$ be a surface geometry for a genus-one phase of $\fT_{1,S^1}^{\rho_1}$, which means that $\cG_1$ comes with a preferred choice of $\P^1$ fibers for the Hirzebruch surfaces constituting $\cG_1$. Similarly, let $\cG_2$ be a surface geometry for a genus-one phase of $\fT_{2,S^1}^{\rho_2}$, which means that $\cG_2$ comes with a preferred choice of $\P^1$ fibers for the Hirzebruch surfaces constituting $\cG_2$. Now exchange $e$ and $f$ curves of a subset of degree-0 Hirzebruch surfaces in $\cG_1$, which can be thought of as performing \textit{local S-dualities} at the location of these surfaces \cite{Bhardwaj:2019ngx}. Assume $\cG_1$ is converted to $\cG_2$ after this exchange\footnote{As we will see in examples, one might also need to interpret some of the Hirzebruch surfaces as describing nodes with trivial $\fh_i$.}. Then, $\cG_1$ and $\cG_2$ are isomorphic as surface geometries, that is they describe the same surface geometry once we forget the special choice of $\P^1$ fibers, and we have found that $\fT_{1,S^1}^{\rho_1}$ is T-dual to $\fT_{2,S^1}^{\rho_2}$.

\section{Examples of T-dualities}
\subsection{Gauge Rank 1}
At gauge rank 1 in 6d, we have three LSTs 
\bit
\item $\su(2)$ with 16 hypers in fundamental. We represent this theory by the graph
\be
\begin{tikzpicture} [scale=1.9]
\node at (-4.6,0.8) {0};
\node at (-4.6,1.1) {$\su(2)$};
\end{tikzpicture}
\ee
\item $\su(2)$ with a hyper in adjoint and discrete theta angle 0. We represent this theory by the graph
\be
\begin{tikzpicture} [scale=1.9]
\node at (-4.6,0.8) {$0_1$};
\node at (-4.6,1.1) {$\su(2)_0$};
\end{tikzpicture}
\ee
where the subscript 1 is placed to distinguish it from the graph associated to the previous LST, and indicates that the base curve used in F-theory construction of the LST has genus 1 (while the base curve for previous LST has genus 0). This theory actually has $\cN=(1,1)$ supersymmetry.
\item $\su(2)$ with a hyper in adjoint and discrete theta angle $\pi$. We represent this theory by the graph
\be
\begin{tikzpicture} [scale=1.9]
\node at (-4.6,0.8) {$0_1$};
\node at (-4.6,1.1) {$\su(2)_\pi$};
\end{tikzpicture}
\ee
This theory also has $\cN=(1,1)$ supersymmetry.
\eit
Only untwisted circle compactification is possible for all three LSTs.

For the untwisted circle compactification of the first LST, a surface geometry is
\be
\begin{tikzpicture} [scale=1.9]
\node (v1) at (-4.3,-0.5) {$\mathbf{0}^8_0$};
\node (v2) at (-2,-0.5) {$\mathbf{1}^{8}_{0}$};
\draw  (v1) edge (v2);
\node at (-3.7,-0.4) {\scriptsize{$2e$+$2f$-$\sum x_i$}};
\node at (-2.6,-0.4) {\scriptsize{$2e$+$2f$-$\sum x_i$}};
\end{tikzpicture}
\ee
Each surface can also be interpreted as a surface associated to a node with $\Omega=1$ and trivial gauge algebra. Thus, we have the T-duality
\be
\begin{tikzpicture} [scale=1.9]
\node at (-4.9,0.9) {$=$};
\node (v1) at (-4.2,0.8) {1};
\node at (-4.2,1.1) {$\sp(0)^{(1)}$};
\draw  (-6.2,1.4) rectangle (-2.6,0.5);
\node (v2) at (-3.2,0.8) {1};
\node at (-3.2,1.1) {$\sp(0)^{(1)}$};
\draw  (v1) edge (v2);
\begin{scope}[shift={(-1,0)}]
\node at (-4.6,0.8) {0};
\node at (-4.6,1.1) {$\su(2)^{(1)}$};
\end{scope}
\end{tikzpicture}
\ee

For the untwisted circle compactification of the second LST, a surface geometry is
\be
\begin{tikzpicture} [scale=1.9]
\node (v1) at (-1.8,-2) {$\mathbf{0}_{0}^{1+1}$};
\node (v10) at (0.4,-2) {\scriptsize{2}};
\draw  (v10) edge (v1);
\node at (-0.9,-1.9) {\scriptsize{$2e$+$f$-$x$-$2y,f$-$x$}};
\draw (v1) .. controls (-2.4,-2.7) and (-1.2,-2.7) .. (v1);
\node at (-2.2,-2.3) {\scriptsize{$x$}};
\node at (-1.4,-2.3) {\scriptsize{$y$}};
\begin{scope}[shift={(4.4,0)}]
\node (v1) at (-1.8,-2) {$\mathbf{1}_{0}^{1+1}$};
\draw (v1) .. controls (-2.4,-2.7) and (-1.2,-2.7) .. (v1);
\node at (-2.2,-2.3) {\scriptsize{$x$}};
\node at (-1.4,-2.3) {\scriptsize{$y$}};
\end{scope}
\draw  (v10) edge (v1);
\node at (1.7,-1.9) {\scriptsize{$2e$+$f$-$x$-$2y,f$-$x$}};
\end{tikzpicture}
\ee
Each surface can also be interpreted as a surface associated to a node with $\Omega=2$ and trivial gauge algebra. Thus, we have the T-duality
\be
\begin{tikzpicture} [scale=1.9]
\node at (-4.9,0.9) {$=$};
\node (v1) at (-4.2,0.8) {2};
\node at (-4.2,1.1) {$\su(1)^{(1)}$};
\draw  (-6.2,1.4) rectangle (-2.6,0.5);
\node (v2) at (-3.2,0.8) {2};
\node at (-3.2,1.1) {$\su(1)^{(1)}$};
\begin{scope}[shift={(-1,0)}]
\node at (-4.6,0.8) {$0_1$};
\node at (-4.6,1.1) {$\su(2)_0^{(1)}$};
\end{scope}
\node (v3) at (-3.7,0.8) {\tiny{2}};
\draw  (v1) edge (v3);
\draw  (v3) edge (v2);
\end{tikzpicture}
\ee
where we have a double edge because the gluing rules imply that $2e+2f-2x-2y$ in $S_0$ is glued to $2e+2f-2x-2y$ in $S_1$, both of which are twice the elliptic fibers $e+f-x-y$. The right hand side denotes the untwisted circle compactification of the $A_1$ $\cN=(2,0)$ LST.

For the untwisted circle compactification of the third LST, a surface geometry is
\be
\begin{tikzpicture} [scale=1.9]
\node (v1) at (-1.8,-2) {$\mathbf{0}_{0}^{1+1}$};
\node (v10) at (0.4,-2) {\scriptsize{2}};
\draw  (v10) edge (v1);
\node at (-0.9,-1.9) {\scriptsize{$2e$+$f$-$x$-$2y,f$-$x$}};
\draw (v1) .. controls (-2.4,-2.7) and (-1.2,-2.7) .. (v1);
\node at (-2.2,-2.3) {\scriptsize{$x$}};
\node at (-1.4,-2.3) {\scriptsize{$y$}};
\begin{scope}[shift={(4.4,0)}]
\node (v1) at (-1.8,-2) {$\mathbf{1}_{1}^{1+1}$};
\draw (v1) .. controls (-2.4,-2.7) and (-1.2,-2.7) .. (v1);
\node at (-2.2,-2.3) {\scriptsize{$x$}};
\node at (-1.4,-2.3) {\scriptsize{$y$}};
\end{scope}
\draw  (v10) edge (v1);
\node at (1.8,-1.9) {\scriptsize{$2h$-$x$-$2y,f$-$x$}};
\end{tikzpicture}
\ee
The surface $S_0$ can also be interpreted as a surface associated to a node with $\Omega=2$ and trivial gauge algebra, while the surface $S_1$ can also be interpreted as a surface associated to a node with value 2 (but $\Omega=1$) and a loop along with trivial gauge algebra. Thus, we have the T-duality
\be
\begin{tikzpicture} [scale=1.9]
\node at (-4.9,0.9) {$=$};
\node (v1) at (-4.2,0.8) {2};
\node at (-4.2,1.1) {$\su(1)^{(1)}$};
\draw  (-6.2,1.4) rectangle (-2.6,0.2);
\node (v2) at (-3.2,0.8) {2};
\node at (-3.2,1.1) {$\su(1)^{(1)}$};
\begin{scope}[shift={(-1,0)}]
\node at (-4.6,0.8) {$0_1$};
\node at (-4.6,1.1) {$\su(2)_\pi^{(1)}$};
\end{scope}
\node (v3) at (-3.7,0.8) {\tiny{2}};
\draw[<-]  (v1) edge (v3);
\draw  (v3) edge (v2);
\draw (v2) .. controls (-3.6,0.3) and (-2.8,0.3) .. (v2);
\end{tikzpicture}
\ee
where we have a directed edge with 2 in the middle because the gluing rules imply that twice the elliptic fiber $2e+2f-2x-2y$ of $S_0$ is glued to the genus-one fiber $2h+f-2x-2y$ of $S_1$. On the right hand side of the above equality, we have a twisted circle compactification of the $D_5$ $\cN=(2,0)$ LST. The twist is performed using a $\Z_4$ 0-form symmetry of the $D_5$ $\cN=(2,0)$ LST whose action on the nodes of the affine $\so(10)^{(1)}$ algebra is such that the affine node goes to spinor node, the spinor node goes to vector node, the vector node goes to cospinor node, the cospinor node goes to affine node, and the other two remaining (trivalent) nodes are exchanged with each other.

\subsection{Gauge Rank 2}
At gauge rank 2 in 6d, we have the following LSTs and their possible circle compactifications
\bit
\item $\su(3)$ with 18 hypers in fundamental. We represent this theory by the graph
\be

\ee
This theory can have twisted compactification by the $\Z_2$ outer-automorphism.
\item $\su(3)$ with a hyper in 2-index symmetric and a hyper in fundamental. We represent this theory by the graph
\be
%
\ee
This theory can have twisted compactification by the $\Z_2$ outer-automorphism but we do not discuss its T-dual.
\item $\su(3)$ with a hyper in adjoint and rational theta angle $2\pi p/q$. We represent this theory by the graph
\be
%
\ee
where the subscript 1 indicates that the base curve used in F-theory construction of the LST has genus 1. This theory actually has $\cN=(1,1)$ supersymmetry.\\
We only discuss T-duals for untwisted compactification of the $\theta=0$ LST in the general rank subsection. A compactification twisted by $\Z_2$ outer-automorphism is possible only for $\theta=0,\pi$, and we discuss both of these cases in this subsection.
\item $\sp(2)$ with a hyper in antisymmetric and 16 hypers in fundamental. We represent this theory by the graph
\be
%
\ee
We will discuss two T-duals for the untwisted compactification of this LST. We discuss one in this subsection, and the other in the subsection on general gauge rank.
\item $\fg_2$ with 10 hypers in irrep of dimension 7. We represent this theory by the graph
\be
%
\ee
\eit

First of all, consider the surface geometry
\be
%
\ee
which describes the untwisted compactification of the $\su(3)$ theory with 18 fundamentals. Exchanging the $\P^1$ fibers $e$ and $f$ in the surfaces $S_1$ and $S_0$, we can rewrite the surface geometry as
\be
%
\ee
The surfaces $S_0$ and $S_1$ now describe an $\sp(1)^{(1)}$ and the surface $S_2$ can be recognized to be describing $\sp(0)^{(1)}$, leading to the T-duality
\be
%
\ee

Now consider another surface geometry for the untwisted compactification of the $\su(3)+18\F$ theory
\be
%
\ee
Now exchanging $e$ and $f$ $\P^1$ fibers in surfaces $S_0$ and $S_1$ leads to the surface geometry
\be
%
\ee
The three surfaces now describe an $\sp(2)^{(1)}$, leading to the T-duality
\be
%
\ee
where the right hand side is the untwisted compactification of the $\sp(2)+\L^2+16\F$ LST.

Moving onto twisted compactifications of the $\su(3)+18\F$ LST, we have the following surface geometry
\be
%
\ee
The surface $S_0$ can be recognized as a surface for $\su(1)^{(1)}$ and the surface $S_1$ can be recognized as surface for $\sp(0)^{(1)}$. From the gluing of the elliptic fibers, we read the precise edge between the two nodes, leading to the T-duality
\be
%
\ee

For the twisted compactification of the $\su(3)_0+\A$ $\cN=(1,1)$ LST, we have the following surface geometry
\be
%
\ee
We can interpret both $S_0$ and $S_1$ as describing $\su(2)^{(1)}$, leading to the T-duality
\be
%
\ee
The right hand side is a circle compactification of $D_4$ $\cN=(2,0)$ LST twisted by the $\Z_4$ symmetry rotating the affine, spinor, cospinor and vector nodes of $\so(8)^{(1)}$ Dynkin diagram into each other.

For the twisted compactification of the $\su(3)_\pi+\A$ $\cN=(1,1)$ LST, we have the following surface geometry
\be
%
\ee
In a similar way as above, we are led to the T-duality
\be
%
\ee
The right hand side is a circle compactification of $A_2$ $\cN=(2,0)$ LST twisted by the $\Z_2$ symmetry exchanging two adjacent nodes of the $\su(3)^{(1)}$ Dynkin diagram.

Now, consider the following surface geometry
\be
%
\ee
which describes untwisted circle compactification of $\su(3)+\S^2+\F$ LST. Exchanging $e$ and $f$ $\P^1$ fibers of surfaces $S_0$ and $S_1$ leads to the T-duality
\be
%
\ee

Finally, consider the surface geometry
\be
%
\ee
As written, the three surfaces describe untwisted circle compactification of the LST $\fg_2+10\F$. Exchanging $e$ and $f$ fibers in $S_1$ converts it into a surface geometry such that the three surfaces describe compactification of the LST $\su(4)+2\L^2+16\F$ twisted by the $\Z_2$ outer automorphism, leading to the T-duality
\be
%
\ee

\subsection{Gauge Rank 3}
Now we discuss some T-dualities that appear at gauge rank 3, but that do not fall into the general cases discussed later.

Consider the surface geometry
\be
%
\ee
which as written describe the compactification of the LST $\su(4)+2\L^2+16\F$ twisted by the $\Z_2$ outer automorphism. Equivalently, the surfaces $S_0$ and $S_2$ can be interpreted as describing $\su(1)^{(1)}$ and the surface $S_1$ can be interpreted as describing $\sp(0)^{(1)}$, leading to the T-duality
\be
%
\ee

The surface geometry
\be
%
\ee
describes the $\Z_2$ outer-automorphism twisted compactification of the $\su(4)_0+\A$ $\cN=(1,1)$ LST. Each surface can be interpreted as describing $\su(1)^{(1)}$, which leads to a T-duality with compactification of $A_{3}$ $\cN=(2,0)$ LST twisted by $\Z_2$ symmetry exchanging the fundamental and anti-fundamental nodes, while leaving the other two nodes invariant of the Dynkin diagram of $\su(4)^{(1)}$
\be
%
\ee
The various $\su(1)^{(1)}$ on the right hand side form the Dynkin diagram of $\sp(2)^{(1)}$. Note that $\su(4)^{(2)}$ and $\sp(2)^{(1)}$ are Langlands dual, i.e. their Dynkin diagrams are related by reversing the directions of arrows.

The surface geometry
\be
%
\ee
describes the $\Z_2$ outer-automorphism twisted compactification of the $\su(4)_\pi+\A$ $\cN=(1,1)$ LST. Each surface can be interpreted as describing $\su(1)^{(1)}$, which leads to a T-duality with compactification of $D_{5}$ $\cN=(2,0)$ LST twisted by $\Z_2$ symmetry exchanging the affine and spinor nodes, the vector and cospinor nodes, and the two trivalent nodes of the Dynkin diagram of $\so(10)^{(1)}$
\be
%
\ee

The surface geometry
\be
%
\ee
describes untwisted circle compactification of the LST $\sp(3)+\L^2+16\F$. Exchanging $e$ and $f$ on surface $S_1$ leads to the following T-duality
\be
%
\ee

The surface geometry
\be
%
\ee
describes untwisted circle compactification of the LST $\sp(3)+\half\L^3+\frac{35}2\F$. Exchanging $e$ and $f$ on surface $S_1$ leads to the following T-duality
\be
%
\ee

\subsection{LSTs with $\so$ gauge algebra}
\subsubsection{Twisted Compactifications}
The surface geometry
\be
%
\ee
describes compactification of LST $\so(8)+4\F+4\S+4\C$ with a $\Z_3$ outer-automorphism twist. Exchanging $e$ and $f$ on $S_1$ leads to the T-duality
\be
%
\ee
where $\su(3)^{(2)}$ is formed by the surfaces $S_0$ and $S_1$.

The surface geometry
\be
%
\ee
describes compactification of LST $\so(8)+4\F+4\S+4\C$ with a $\Z_2$ outer-automorphism twist. Exchanging $e$ and $f$ on $S_1$ leads to the T-duality
\be
%
\ee
where $\su(3)^{(2)}$ is formed by the surfaces $S_0$ and $S_1$.

The surface geometry
\be
%
\ee
describes compactification of LST $\so(10)+6\F+4\S$ with a $\Z_2$ outer-automorphism twist. Exchanging $e$ and $f$ on $S_2$ leads to the T-duality
\be
%
\ee
where $\su(5)^{(2)}$ is formed by the surfaces $S_2$, $S_3$ and $S_4$.

The surface geometry
\be
%
\ee
describes compactification of LST $\so(12)+8\F+\S+\C$ with a $\Z_2$ outer-automorphism twist. Exchanging $e$ and $f$ on $S_2$ leads to the T-duality
\be
%
\ee
where $\su(7)^{(2)}$ is formed by the surfaces $S_2$, $S_3$, $S_4$ and $S_5$.

The surface geometry
\be
%
\ee
describes compactification of LST $\so(14)+10\F+\S$ with a $\Z_2$ outer-automorphism twist. Exchanging $e$ and $f$ on $S_2$ leads to the T-duality
\be
%
\ee
where $\su(9)^{(2)}$ is formed by the surfaces $S_2$, $S_3$, $S_4$, $S_5$ and $S_6$.

\subsubsection{Untwisted Compactifications}
The surface geometry
\be
%
\ee
describes untwisted compactification of LST $\so(7)+3\F+8\S$. Exchanging $e$ and $f$ on $S_0$ and $S_1$ leads to the T-duality
\be
%
\ee
where $\su(2)^{(1)}$ is formed by the surfaces $S_0$ and $S_1$.

Similarly, the surface geometry
\be
%
\ee
describes untwisted compactification of LST $\so(9)+5\F+4\S$. Exchanging $e$ and $f$ on $S_2$ leads to the T-duality
\be
%
\ee
where $\su(4)^{(2)}$ is formed by the surfaces $S_0$, $S_1$ and $S_2$.

In a similar fashion, we can deduce the following T-dualities
\be
%

The surface geometry
\be
%
\ee
describes untwisted compactification of LST $\so(8)+4\F+4\S+4\C$. Exchanging $e$ and $f$ on $S_0$, $S_1$, $S_3$ and $S_4$ leads to the T-duality
\be
%
\ee
where one $\su(2)^{(1)}$ is formed by the surfaces $S_0$ and $S_1$, and the other $\su(2)^{(1)}$ is formed by the surfaces $S_3$ and $S_4$.

The surface geometry
\be
%
\ee
describes untwisted compactification of LST $\so(10)+6\F+4\S$. Exchanging $e$ and $f$ on $S_2$, $S_4$ and $S_5$ leads to the T-duality
\be
%
\ee
where $\su(4)^{(2)}$ is formed by the surfaces $S_0$, $S_1$ and $S_2$, and $\su(2)^{(1)}$ is formed by the surfaces $S_4$ and $S_5$.

The surface geometry
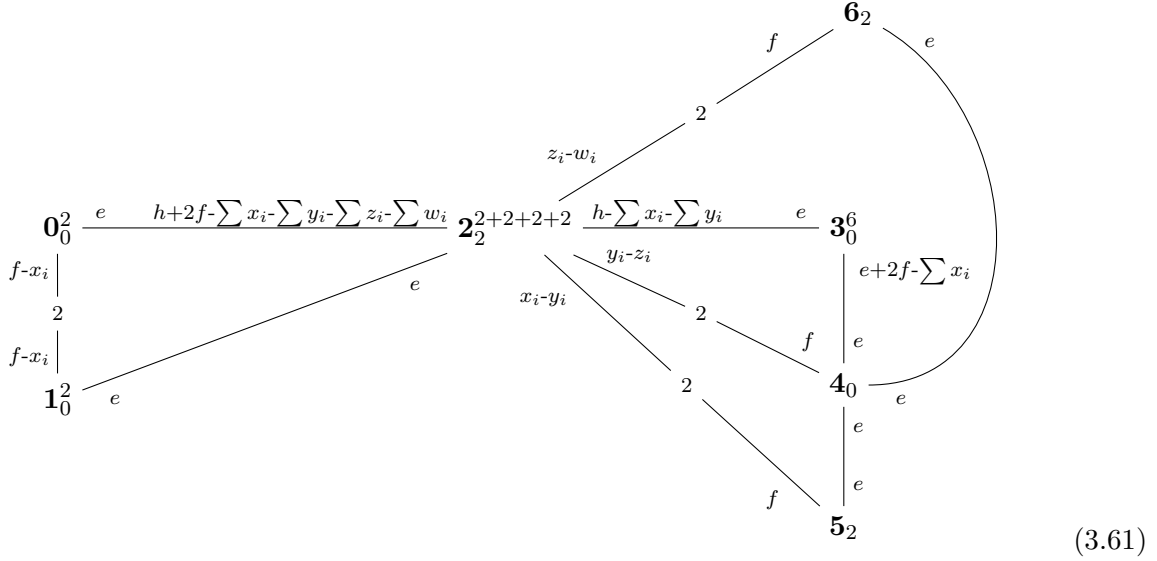
\be
%
\ee
describes untwisted compactification of LST $\so(12)+8\F+2\S$. Exchanging $e$ and $f$ on $S_0$, $S_1$ and $S_3$ leads to the T-duality
\be
%
\ee
where $\su(6)^{(2)}$ is formed by the surfaces $S_3$, $S_4$, $S_5$ and $S_6$, and $\su(2)^{(1)}$ is formed by the surfaces $S_0$ and $S_1$. The subscript on $\sp(0)$ denotes that theta angle is 0, which means that the flavor symmetry attached to the node is $\u(2)$.

The surface geometry
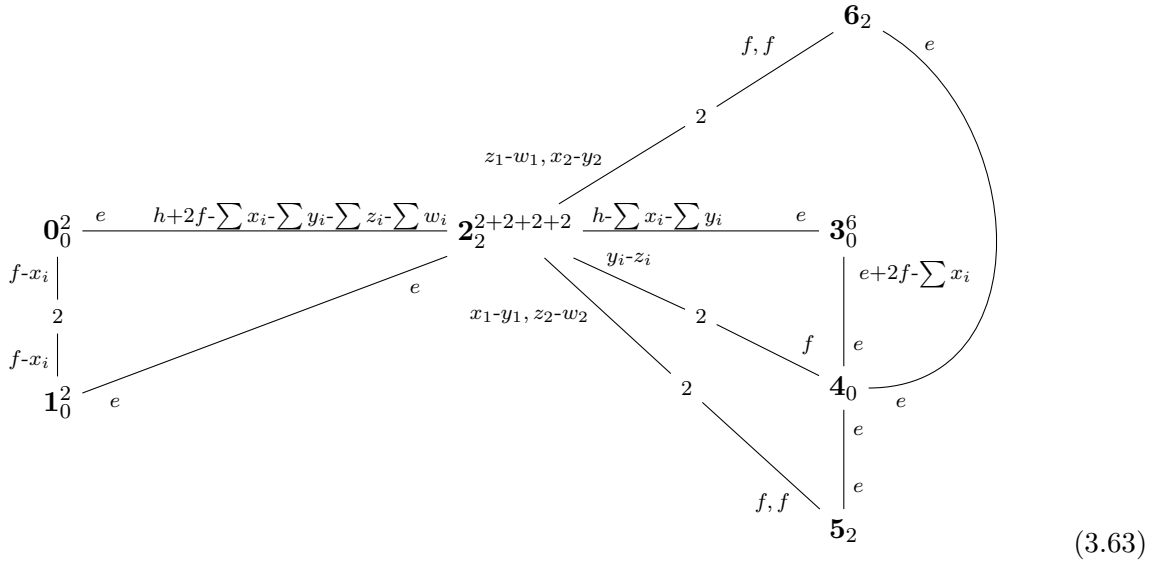
\be
%
\ee
describes untwisted compactification of LST $\so(12)+8\F+\S+\C$. Exchanging $e$ and $f$ on $S_0$, $S_1$ and $S_3$ leads to the T-duality
\be
%
\ee
where $\su(6)^{(2)}$ is formed by the surfaces $S_3$, $S_4$, $S_5$ and $S_6$, and $\su(2)^{(1)}$ is formed by the surfaces $S_0$ and $S_1$. The subscript on $\sp(0)$ denotes that theta angle is $\pi$, which means that the flavor symmetry attached to the node is $\u(1)\oplus\u(1)$.

The surface geometry
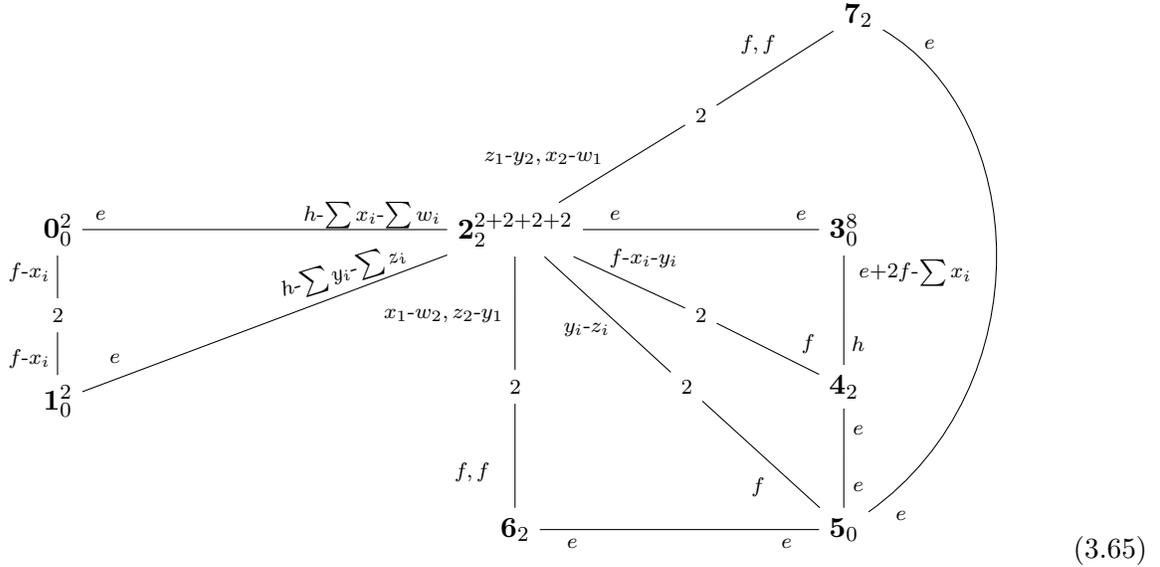
\be
%
\ee
describes untwisted compactification of LST $\so(14)+10\F+\S$. Exchanging $e$ and $f$ on $S_0$, $S_1$ and $S_3$ leads to the T-duality
\be
%
\ee
where $\su(8)^{(2)}$ is formed by the surfaces $S_3$, $S_4$, $S_5$, $S_6$ and $S_7$, and $\su(2)^{(1)}$ is formed by the surfaces $S_0$ and $S_1$. The theta angle for $\sp(0)$ must be 0, because $\sp(0)_\pi$ cannot couple to $\su(8)\oplus\su(2)$ on neighboring nodes.

\subsection{LSTs with $\su(6)$ gauge algebra}
In this subsection, we discuss T-dualities of LSTs with $\su(6)$ gauge algebra which do not fall into general classes discussed later.

The surface geometry
\be
%
\ee
describes untwisted compactification of LST $\su(6)+16\F+2\L^2$. Exchanging $e$ and $f$ on $S_5$ leads to the T-duality
\be
%
\ee
where the theta angle for $\sp(0)$ is 0, because it allows coupling to an $\su(2)$ flavor symmetry which is shown in square brackets. The gluing curves in $S_0$ to the $\P^1$ fibers of the $\P^1$-fibered non-compact surfaces generating the $\su(2)^{(1)}$ corresponding to this flavor symmetry are $h-x_2-x_3-x_4-y_1$ and $h-y_2-y_3-y_4-x_1$. One can understand the above as first coupling $\sp(0)_0$ to $\su(8)\oplus\su(2)$ and then decomposing $\su(8)$ to $\sp(4)$.

The surface geometry
\be
%
\ee
describes untwisted compactification of LST $\su(6)+17\F+\L^2+\half\L^3$. Exchanging $e$ and $f$ on $S_5$ leads to the T-duality
\be
%
\ee
where the theta angle for $\sp(0)$ is $\pi$, because it allows coupling to a $\u(1)$ flavor symmetry which is shown in square brackets. One can understand the above as first coupling $\sp(0)_\pi$ to $\su(8)\oplus\u(1)$ and then decomposing $\su(8)$ to $\sp(4)$.

The surface geometry
\be
%
\ee
describes untwisted compactification of LST $\su(6)+18\F+\L^3$. Exchanging $e$ and $f$ on $S_5$ leads to the T-duality
\be\label{new}
%
\ee
where the theta angle for $\sp(0)$ is neither 0 nor $\pi$ and hence is denoted by the subscript`new'. It allows coupling to an $\su(3)$ flavor symmetry which is shown in square brackets. The gluing curves in $S_0$ to the $\P^1$ fibers of the $\P^1$-fibered non-compact surfaces generating the $\su(3)^{(1)}$ corresponding to this flavor symmetry are $h-x_1-y_2-z_2-w_1$, $h-x_2-y_1-z_1-w_1$ and $w_1-w_2$. One can understand the above as first coupling $\sp(0)$ to $\fe_6\oplus\su(3)$ and then decomposing $\fe_6$ to $\sp(4)$.

The T-duals for compactifications twisted by $\Z_2$ outer-automorphisms of the above three theories are as follows. The surface geometry 
\be

\ee
describes twisted compactification of LST $\su(6)+16\F+2\L^2$. Exchanging $e$ and $f$ on $S_0$, $S_1$ and $S_2$ leads to the T-duality
\be
%
\ee
where $\su(2)^{(1)}$ is formed by the surfaces $S_0$ and $S_1$, and $\sp(1)^{(1)}$ is formed by the surfaces $S_2$ and $S_3$.

The surface geometry
\be
%
\ee
describes twisted compactification of LST $\su(6)+17\F+\half\L^3+\L^2$. Exchanging $e$ and $f$ on $S_0$ and $S_1$ leads to the T-duality
\be
%
\ee
where $\su(2)^{(1)}$ is formed by the surfaces $S_0$ and $S_1$.

The surface geometry
\be
%
\ee
describes twisted compactification of LST $\su(6)+18\F+\L^3$. Exchanging $e$ and $f$ on $S_2$ leads to the T-duality
\be
%
\ee
where $\sp(1)^{(1)}$ is formed by the surfaces $S_2$ and $S_3$.

\subsection{General Gauge Rank}
\subsubsection{$\cN=(1,0)$ Examples}
The surface geometry
\be
%
\ee
can be interpreted as describing untwisted compactification of $\sp(m+1)+\L^2+16\F$ LST. Equivalently, the surfaces $S_0$ and $S_{m+1}$ can be interpreted as describing $\sp(0)^{(1)}$ and the other surfaces describing $\su(1)^{(1)}$, leading to the T-duality
\be
%
\ee

The surface geometry
\be
%
\ee
can be interpreted as describing $\Z_2$ outer-automorphism twisted compactification of $\su(2m+1)+2\L^2+16\F$ LST. Equivalently, the surfaces $S_0$ can be interpreted as describing $\sp(0)^{(1)}$ and the other surfaces describing $\su(1)^{(1)}$, leading to the T-duality
\be
%
\ee

The surface geometry
\be
%
\ee
can be interpreted as describing $\Z_2$ outer-automorphism twisted compactification of $\su(2m)+2\L^2+16\F$ LST. Equivalently, the surfaces $S_0$ can be interpreted as describing $\sp(0)^{(1)}$ and the other surfaces describing $\su(1)^{(1)}$, leading to the T-duality
\be
%
\ee

The surface geometry
\be
%
\ee
describes untwisted compactification of LST $\su(2m+1)+\S^2+\L^2$. Exchanging $e$ and $f$ on all surfaces except $S_{m+1}$ leads to the T-duality
\be
%
\ee
where various $\su(2)^{(1)}$ without a loop are formed by pairs of surfaces $(S_i,S_{2m+2-i})$ for $2\le i\le m$, and $(S_0,S_1)$ forms $\su(2)^{(1)}$ with loop.

The surface geometry
\be
%
\ee
describes untwisted compactification of LST $\su(2m)+\S^2+\L^2$. Exchanging $e$ and $f$ on all surfaces leads to the T-duality
\be
%
\ee
where various $\su(2)^{(1)}$ without a loop are formed by pairs of surfaces $(S_i,S_{2m+1-i})$ for $2\le i\le m-1$, and $(S_0,S_1)$ and $(S_m,S_{m+1})$ form $\su(2)^{(1)}$ with loop.

The surface geometry
\be
%
\ee
describes untwisted compactification of LST $\su(2m+2)+16\F+2\L^2$. Exchanging $e$ and $f$ on all surfaces except $S_0$ and $S_{m+1}$ leads to the T-duality
\be
%
\ee
where various $\su(2)^{(1)}$ are formed by pairs of surfaces $(S_i,S_{2m+2-i})$ for $1\le i\le m$.

The surface geometry
\be
%
\ee
describes untwisted compactification of LST $\su(2m+1)+16\F+2\L^2$. Exchanging $e$ and $f$ on all surfaces except $S_{m+1}$ leads to the T-duality
\be
%
\ee
where various $\su(2)^{(1)}$ are formed by pairs of surfaces $(S_i,S_{2m+2-i})$ for $2\le i\le m$, and $\sp(1)^{(1)}$ is formed by $S_0$ and $S_1$.

The surface geometry
\be
%
\ee
describes untwisted compactification of LST $\su(2m)+16\F+2\L^2$. Exchanging $e$ and $f$ on $S_0$ and $S_1$ leads to a description of the untwisted compactification of LST $\sp(2m-1)+16\F+\L^2$. If we also exchange $e$ and $f$ on $S_m$ and $S_{m+1}$, then we are lead to a surface geometry in which surfaces $S_1$ to $S_m$ describe an $\sp(m-1)^{(1)}$, and surfaces $S_{m+1}$ to $S_{2m-1}$ and $S_0$ describe an $\sp(m-1)^{(1)}$. In total, we have the following T-dualities
\be
%
\ee

Similarly, the surface geometry
\be
%
\ee
describes untwisted compactification of LST $\su(2m+1)+16\F+2\L^2$. Exchanging $e$ and $f$ on $S_0$ and $S_1$ leads to a description of the untwisted compactification of LST $\sp(2m)+16\F+\L^2$. If we also exchange $e$ and $f$ on $S_m$ and $S_{m+1}$, then we are lead to a surface geometry in which surfaces $S_1$ to $S_m$ describe an $\sp(m-1)^{(1)}$, and surfaces $S_{m+1}$ to $S_{2m}$ and $S_0$ describe an $\sp(m)^{(1)}$. In total, we have the following T-dualities.
\be
%
\ee

\subsubsection{$\cN=(1,1)$ and $\cN=(2,0)$ LSTs}
For untwisted compactification of $\su(m)$ $\cN=(1,1)$ LST with $\theta=0$, we have the surface geometry
\be
%
\ee
in which each surface can be interpreted as describing an $\su(1)^{(1)}$, which leads to a T-duality with untwisted compactification of $A_{m-1}$ $\cN=(2,0)$ LST
\be
%
\ee

For $\Z_2$ outer-automorphism twisted compactification of $\su(2m)$ $\cN=(1,1)$ LST with $\theta=0$, we have the surface geometry
\be
%
\ee
in which each surface can be interpreted as describing an $\su(1)^{(1)}$, which leads to a T-duality with compactification of $D_{m+1}$ $\cN=(2,0)$ LST twisted by $\Z_2$ symmetry exchanging the spinor and cospinor nodes
\be
%
\ee
where the various $\su(1)^{(1)}$ on the right hand side form the Dynkin diagram of $\so(2m+1)^{(1)}$. Note that $\su(2m)^{(2)}$ and $\so(2m+1)^{(1)}$ are Langlands dual, i.e. their Dynkin diagrams are related by reversing the directions of arrows.

For $\Z_2$ outer-automorphism twisted compactification of $\su(2m)$ $\cN=(1,1)$ LST with $\theta=\pi$, we have the surface geometry
\be
%
\ee
in which each surface can be interpreted as describing an $\su(1)^{(1)}$, which leads to a T-duality with compactification of $D_{2m+1}$ $\cN=(2,0)$ LST twisted by $\Z_2$ symmetry exchanging the spinor node with affine node and cospinor node with vector node (the action on other nodes is fixed by these requirements)
\be
%
\ee

For $\Z_2$ outer-automorphism twisted compactification of $\su(2m+1)$ $\cN=(1,1)$ LST with $\theta=0$, we have the surface geometry
\be
%
\ee
in which each surface can be interpreted as describing an $\su(1)^{(1)}$, which leads to a T-duality with compactification of $D_{2m+2}$ $\cN=(2,0)$ LST twisted by $\Z_4$ symmetry that sends affine node to spinor node, spinor node to vector node, vector node to cospinor node, and cospinor node to vector node
\be
%
\ee

For $\Z_2$ outer-automorphism twisted compactification of $\su(2m+1)$ $\cN=(1,1)$ LST with $\theta=\pi$, we have the surface geometry
\be
%
\ee
in which each surface can be interpreted as describing an $\su(1)^{(1)}$, which leads to a T-duality with compactification of $A_{2m}$ $\cN=(2,0)$ LST twisted by $\Z_2$ symmetry that leaves affine node invariant but acts on other nodes as $\Z_2$ outer-automorphism of $\su(2m+1)$
\be
%
\ee

For untwisted compactification of $\so(2m+1)$ $\cN=(1,1)$ LST, we have the surface geometry
\be
%
\ee
in which each surface can be interpreted as describing an $\su(1)^{(1)}$, which leads to a T-duality with compactification of $D_{2m}$ $\cN=(2,0)$ LST twisted by $\Z_2$ symmetry exchanging the spinor node with affine node and cospinor node with vector node
\be
%
\ee
The various $\su(1)^{(1)}$ on the right hand side form the Dynkin diagram of $\su(2m)^{(2)}$. Note that $\so(2m+1)^{(1)}$ and $\su(2m)^{(2)}$ are Langlands dual, i.e. their Dynkin diagrams are related by reversing the directions of arrows.

For untwisted compactification of $\sp(m)$ $\cN=(1,1)$ LST with $\theta=0$, we have the surface geometry
\be
%
\ee
in which each surface can be interpreted as describing an $\su(1)^{(1)}$, which leads to a T-duality with compactification of $D_{m+2}$ $\cN=(2,0)$ LST twisted by $\Z_2$ symmetry that sends affine node to vector node and spinor node to cospinor node
\be
%
\ee
The various $\su(1)^{(1)}$ on the right hand side form the Dynkin diagram of $\so(2m+2)^{(2)}$. Note that $\sp(m)^{(1)}$ and $\so(2m+2)^{(2)}$ are Langlands dual, i.e. their Dynkin diagrams are related by reversing the directions of arrows.

For untwisted compactification of $\sp(m)$ $\cN=(1,1)$ LST with $\theta=\pi$, we have the surface geometry
\be
%
\ee
in which each surface can be interpreted as describing an $\su(1)^{(1)}$, which leads to a T-duality with compactification of $D_{2m+3}$ $\cN=(2,0)$ LST twisted by $\Z_4$ symmetry that sends affine node to spinor node, spinor node to vector node, vector node to cospinor node, and cospinor node to vector node
\be
%
\ee

For untwisted compactification of $\so(2m)$ $\cN=(1,1)$ LST, we have the surface geometry
\be
%
\ee
in which each surface can be interpreted as describing an $\su(1)^{(1)}$, which leads to a T-duality with untwisted compactification of $D_{m}$ $\cN=(2,0)$ LST
\be
%
\ee

For $\Z_2$ outer-automorphism twisted compactification of $\so(2m+2)$ $\cN=(1,1)$ LST, we have the surface geometry
\be
%
\ee
in which each surface can be interpreted as describing an $\su(1)^{(1)}$, which leads to a T-duality with compactification of $A_{2m-1}$ $\cN=(2,0)$ LST twisted by $\Z_2$ symmetry that leaves affine node invariant but acts on other nodes as $\Z_2$ outer-automorphism of $\su(2m)$
\be
%
\ee
The various $\su(1)^{(1)}$ on the right hand side form the Dynkin diagram of $\sp(m)^{(1)}$. Note that $\so(2m+2)^{(2)}$ and $\sp(m)^{(1)}$ are Langlands dual, i.e. their Dynkin diagrams are related by reversing the directions of arrows.

\subsection{Other T-dualities between $\cN=(1,1)$ and $\cN=(2,0)$}
In a similar way, the reader can derive remaining T-dualities of compactifications of $\cN=(1,1)$ LSTs, for which we quote the results below:
\bit
\item The untwisted compactification of $\fe_n$ $\cN=(1,1)$ LST is T-dual to untwisted compactification of $E_n$ $\cN=(2,0)$ LST, for each $6\le n\le 8$.
\item The $\Z_2$ outer-automorphism twisted compactification of $\fe_6$ $\cN=(1,1)$ LST is T-dual to compactification of $E_6$ $\cN=(2,0)$ LST twisted by a $\Z_2$ symmetry exchanging two of the three single valent nodes. The $\su(1)^{(1)}$ involved in the compactification of the $\cN=(2,0)$ theory arrange themselves in the form of Dynkin diagram of $\ff_4^{(1)}$ which is Langlands dual to the affine algebra $\fe_6^{(2)}$ related to the compactification of the T-dual $\cN=(1,1)$ LST.
\item The untwisted compactification of $\ff_4$ $\cN=(1,1)$ LST is T-dual to compactification of $E_7$ $\cN=(2,0)$ LST twisted by a $\Z_2$ symmetry exchanging two of the three single valent nodes. The $\su(1)^{(1)}$ involved in the compactification of the $\cN=(2,0)$ theory arrange themselves in the form of Dynkin diagram of $\fe_6^{(2)}$ which is Langlands dual to the affine algebra $\ff_4^{(1)}$ related to the compactification of the T-dual $\cN=(1,1)$ LST.
\item The untwisted compactification of $\fg_2$ $\cN=(1,1)$ LST is T-dual to compactification of $E_6$ $\cN=(2,0)$ LST twisted by a $\Z_3$ symmetry rotating the three single valent nodes. The $\su(1)^{(1)}$ involved in the compactification of the $\cN=(2,0)$ theory arrange themselves in the form of Dynkin diagram of $\so(8)^{(3)}$ which is Langlands dual to the affine algebra $\fg_2^{(1)}$ related to the compactification of the T-dual $\cN=(1,1)$ LST.
\item The $\Z_3$ outer-automorphism twisted compactification of $\so(8)$ $\cN=(1,1)$ LST is T-dual to compactification of $D_4$ $\cN=(2,0)$ LST twisted by a $\Z_3$ symmetry rotating three out of four single valent nodes. The $\su(1)^{(1)}$ involved in the compactification of the $\cN=(2,0)$ theory arrange themselves in the form of Dynkin diagram of $\fg_2^{(1)}$ which is Langlands dual to the affine algebra $\so(8)^{(3)}$ related to the compactification of the T-dual $\cN=(1,1)$ LST.
\eit

\subsection*{Acknowledgements}
The author thanks Stefan Hohenegger, Patrick Jefferson and Kantaro Ohmori for related discussions.
This work is partly supported by ERC grants 682608 and 787185 under the European Union’s Horizon 2020 programme, and partly supported by NSF grant PHY-1719924. This material is also partially supported by a grant from the Simons Foundation and the hospitality of the Aspen Center for Physics.

\bibliographystyle{ytphys}
\small 
\baselineskip=.94\baselineskip
\let\bbb\bibitem\def\bibitem{\itemsep4pt\bbb}
\bibliography{ref}

\providecommand{\href}[2]{#2}\begingroup\raggedright\begin{thebibliography}{10}

\bibitem{Seiberg:1997zk}
N.~Seiberg, ``{New theories in six-dimensions and matrix description of M
  theory on T**5 and T**5 / Z(2)},''
  \href{http://dx.doi.org/10.1016/S0370-2693(97)00805-8}{{\em Phys. Lett. B}
  {\bfseries 408} (1997) 98--104},
  \href{http://arxiv.org/abs/hep-th/9705221}{{\ttfamily arXiv:hep-th/9705221}}.

\bibitem{Intriligator:1997dh}
K.~A. Intriligator, ``{New string theories in six-dimensions via branes at
  orbifold singularities},''
  \href{http://dx.doi.org/10.4310/ATMP.1997.v1.n2.a5}{{\em Adv. Theor. Math.
  Phys.} {\bfseries 1} (1998) 271--282},
  \href{http://arxiv.org/abs/hep-th/9708117}{{\ttfamily arXiv:hep-th/9708117}}.

\bibitem{Intriligator:1999cn}
K.~A. Intriligator, ``{Compactified little string theories and compact moduli
  spaces of vacua},'' \href{http://dx.doi.org/10.1103/PhysRevD.61.106005}{{\em
  Phys. Rev. D} {\bfseries 61} (2000) 106005},
  \href{http://arxiv.org/abs/hep-th/9909219}{{\ttfamily arXiv:hep-th/9909219}}.

\bibitem{Bhardwaj:2015oru}
L.~Bhardwaj, M.~Del~Zotto, J.~J. Heckman, D.~R. Morrison, T.~Rudelius, and
  C.~Vafa, ``{F-theory and the Classification of Little Strings},''
  \href{http://dx.doi.org/10.1103/PhysRevD.93.086002}{{\em Phys. Rev. D}
  {\bfseries 93} no.~8, (2016) 086002},
  \href{http://arxiv.org/abs/1511.05565}{{\ttfamily arXiv:1511.05565
  [hep-th]}}. [Erratum: Phys.Rev.D 100, 029901 (2019)].

\bibitem{Bhardwaj:2019hhd}
L.~Bhardwaj, ``{Revisiting the classifications of 6d SCFTs and LSTs},''
  \href{http://dx.doi.org/10.1007/JHEP03(2020)171}{{\em JHEP} {\bfseries 03}
  (2020) 171}, \href{http://arxiv.org/abs/1903.10503}{{\ttfamily
  arXiv:1903.10503 [hep-th]}}.

\bibitem{Bhardwaj:2018jgp}
L.~Bhardwaj, D.~R. Morrison, Y.~Tachikawa, and A.~Tomasiello, ``{The frozen
  phase of F-theory},'' \href{http://dx.doi.org/10.1007/JHEP08(2018)138}{{\em
  JHEP} {\bfseries 08} (2018) 138},
  \href{http://arxiv.org/abs/1805.09070}{{\ttfamily arXiv:1805.09070
  [hep-th]}}.

\bibitem{Seiberg:1996qx}
N.~Seiberg, ``{Nontrivial fixed points of the renormalization group in
  six-dimensions},''
  \href{http://dx.doi.org/10.1016/S0370-2693(96)01424-4}{{\em Phys. Lett. B}
  {\bfseries 390} (1997) 169--171},
  \href{http://arxiv.org/abs/hep-th/9609161}{{\ttfamily arXiv:hep-th/9609161}}.

\bibitem{Danielsson:1997kt}
U.~H. Danielsson, G.~Ferretti, J.~Kalkkinen, and P.~Stjernberg, ``{Notes on
  supersymmetric gauge theories in five-dimensions and six-dimensions},''
  \href{http://dx.doi.org/10.1016/S0370-2693(97)00645-X}{{\em Phys. Lett. B}
  {\bfseries 405} (1997) 265--270},
  \href{http://arxiv.org/abs/hep-th/9703098}{{\ttfamily arXiv:hep-th/9703098}}.

\bibitem{Bhardwaj:2015xxa}
L.~Bhardwaj, ``{Classification of 6d $ \mathcal{N}=\left(1,0\right) $ gauge
  theories},'' \href{http://dx.doi.org/10.1007/JHEP11(2015)002}{{\em JHEP}
  {\bfseries 11} (2015) 002}, \href{http://arxiv.org/abs/1502.06594}{{\ttfamily
  arXiv:1502.06594 [hep-th]}}.

\bibitem{Hohenegger:2015btj}
S.~Hohenegger, A.~Iqbal, and S.-J. Rey, ``{Instanton-monopole correspondence
  from M-branes on $\mathbb S^1$ and little string theory},''
  \href{http://dx.doi.org/10.1103/PhysRevD.93.066016}{{\em Phys. Rev. D}
  {\bfseries 93} no.~6, (2016) 066016},
  \href{http://arxiv.org/abs/1511.02787}{{\ttfamily arXiv:1511.02787
  [hep-th]}}.

\bibitem{Hohenegger:2016eqy}
S.~Hohenegger, A.~Iqbal, and S.-J. Rey, ``{Self-Duality and Self-Similarity of
  Little String Orbifolds},''
  \href{http://dx.doi.org/10.1103/PhysRevD.94.046006}{{\em Phys. Rev. D}
  {\bfseries 94} no.~4, (2016) 046006},
  \href{http://arxiv.org/abs/1605.02591}{{\ttfamily arXiv:1605.02591
  [hep-th]}}.

\bibitem{Hohenegger:2016yuv}
S.~Hohenegger, A.~Iqbal, and S.-J. Rey, ``{Dual Little Strings from F-Theory
  and Flop Transitions},''
  \href{http://dx.doi.org/10.1007/JHEP07(2017)112}{{\em JHEP} {\bfseries 07}
  (2017) 112}, \href{http://arxiv.org/abs/1610.07916}{{\ttfamily
  arXiv:1610.07916 [hep-th]}}.

\bibitem{Bastian:2017ing}
B.~Bastian, S.~Hohenegger, A.~Iqbal, and S.-J. Rey, ``{Dual little strings and
  their partition functions},''
  \href{http://dx.doi.org/10.1103/PhysRevD.97.106004}{{\em Phys. Rev. D}
  {\bfseries 97} no.~10, (2018) 106004},
  \href{http://arxiv.org/abs/1710.02455}{{\ttfamily arXiv:1710.02455
  [hep-th]}}.

\bibitem{Bastian:2017ary}
B.~Bastian, S.~Hohenegger, A.~Iqbal, and S.-J. Rey, ``{Triality in Little
  String Theories},'' \href{http://dx.doi.org/10.1103/PhysRevD.97.046004}{{\em
  Phys. Rev. D} {\bfseries 97} no.~4, (2018) 046004},
  \href{http://arxiv.org/abs/1711.07921}{{\ttfamily arXiv:1711.07921
  [hep-th]}}.

\bibitem{Bastian:2018dfu}
B.~Bastian, S.~Hohenegger, A.~Iqbal, and S.-J. Rey, ``{Beyond Triality: Dual
  Quiver Gauge Theories and Little String Theories},''
  \href{http://dx.doi.org/10.1007/JHEP11(2018)016}{{\em JHEP} {\bfseries 11}
  (2018) 016}, \href{http://arxiv.org/abs/1807.00186}{{\ttfamily
  arXiv:1807.00186 [hep-th]}}.

\bibitem{Bastian:2018jlf}
B.~Bastian and S.~Hohenegger, ``{Dihedral Symmetries of Gauge Theories from
  Dual Calabi-Yau Threefolds},''
  \href{http://dx.doi.org/10.1103/PhysRevD.99.066013}{{\em Phys. Rev. D}
  {\bfseries 99} no.~6, (2019) 066013},
  \href{http://arxiv.org/abs/1811.03387}{{\ttfamily arXiv:1811.03387
  [hep-th]}}.

\bibitem{Jefferson:2018irk}
P.~Jefferson, S.~Katz, H.-C. Kim, and C.~Vafa, ``{On Geometric Classification
  of 5d SCFTs},'' \href{http://dx.doi.org/10.1007/JHEP04(2018)103}{{\em JHEP}
  {\bfseries 04} (2018) 103}, \href{http://arxiv.org/abs/1801.04036}{{\ttfamily
  arXiv:1801.04036 [hep-th]}}.

\bibitem{Bhardwaj:2018yhy}
L.~Bhardwaj and P.~Jefferson, ``{Classifying $5d$ SCFTs via $6d$ SCFTs: Rank
  one},'' \href{http://dx.doi.org/10.1007/JHEP07(2019)178}{{\em JHEP}
  {\bfseries 07} (2019) 178}, \href{http://arxiv.org/abs/1809.01650}{{\ttfamily
  arXiv:1809.01650 [hep-th]}}. [Addendum: JHEP 01, 153 (2020)].

\bibitem{Bhardwaj:2018vuu}
L.~Bhardwaj and P.~Jefferson, ``{Classifying 5d SCFTs via 6d SCFTs: Arbitrary
  rank},'' \href{http://dx.doi.org/10.1007/JHEP10(2019)282}{{\em JHEP}
  {\bfseries 10} (2019) 282}, \href{http://arxiv.org/abs/1811.10616}{{\ttfamily
  arXiv:1811.10616 [hep-th]}}.

\bibitem{Bhardwaj:2019ngx}
L.~Bhardwaj, ``{Dualities of 5d gauge theories from S-duality},''
  \href{http://dx.doi.org/10.1007/JHEP07(2020)012}{{\em JHEP} {\bfseries 07}
  (2020) 012}, \href{http://arxiv.org/abs/1909.05250}{{\ttfamily
  arXiv:1909.05250 [hep-th]}}.

\bibitem{Bhardwaj:2019jtr}
L.~Bhardwaj, ``{On the classification of 5d SCFTs},''
  \href{http://dx.doi.org/10.1007/JHEP09(2020)007}{{\em JHEP} {\bfseries 09}
  (2020) 007}, \href{http://arxiv.org/abs/1909.09635}{{\ttfamily
  arXiv:1909.09635 [hep-th]}}.

\bibitem{Bhardwaj:2019fzv}
L.~Bhardwaj, P.~Jefferson, H.-C. Kim, H.-C. Tarazi, and C.~Vafa, ``{Twisted
  Circle Compactifications of 6d SCFTs},''
  \href{http://dx.doi.org/10.1007/JHEP12(2020)151}{{\em JHEP} {\bfseries 12}
  (2020) 151}, \href{http://arxiv.org/abs/1909.11666}{{\ttfamily
  arXiv:1909.11666 [hep-th]}}.

\bibitem{Bhardwaj:2019xeg}
L.~Bhardwaj, ``{Do all 5d SCFTs descend from 6d SCFTs?},''
  \href{http://dx.doi.org/10.1007/JHEP04(2021)085}{{\em JHEP} {\bfseries 04}
  (2021) 085}, \href{http://arxiv.org/abs/1912.00025}{{\ttfamily
  arXiv:1912.00025 [hep-th]}}.

\bibitem{Bhardwaj:2020gyu}
L.~Bhardwaj and G.~Zafrir, ``{Classification of 5d $ \mathcal{N} $ = 1 gauge
  theories},'' \href{http://dx.doi.org/10.1007/JHEP12(2020)099}{{\em JHEP}
  {\bfseries 12} (2020) 099}, \href{http://arxiv.org/abs/2003.04333}{{\ttfamily
  arXiv:2003.04333 [hep-th]}}.

\bibitem{Bhardwaj:2020kim}
L.~Bhardwaj, ``{More 5d KK theories},''
  \href{http://dx.doi.org/10.1007/JHEP03(2021)054}{{\em JHEP} {\bfseries 03}
  (2021) 054}, \href{http://arxiv.org/abs/2005.01722}{{\ttfamily
  arXiv:2005.01722 [hep-th]}}.

\bibitem{Bhardwaj:2020phs}
L.~Bhardwaj and S.~Sch\"afer-Nameki, ``{Higher-form symmetries of 6d and 5d
  theories},'' \href{http://dx.doi.org/10.1007/JHEP02(2021)159}{{\em JHEP}
  {\bfseries 02} (2021) 159}, \href{http://arxiv.org/abs/2008.09600}{{\ttfamily
  arXiv:2008.09600 [hep-th]}}.

\bibitem{Bhardwaj:2020ruf}
L.~Bhardwaj, ``{Flavor symmetry of 5d SCFTs. Part I. General setup},''
  \href{http://dx.doi.org/10.1007/JHEP09(2021)186}{{\em JHEP} {\bfseries 09}
  (2021) 186}, \href{http://arxiv.org/abs/2010.13230}{{\ttfamily
  arXiv:2010.13230 [hep-th]}}.

\bibitem{Bhardwaj:2020avz}
L.~Bhardwaj, ``{Flavor symmetry of 5$d$ SCFTs. Part II. Applications},''
  \href{http://dx.doi.org/10.1007/JHEP04(2021)221}{{\em JHEP} {\bfseries 04}
  (2021) 221}, \href{http://arxiv.org/abs/2010.13235}{{\ttfamily
  arXiv:2010.13235 [hep-th]}}.

\bibitem{Apruzzi:2021vcu}
F.~Apruzzi, S.~Schafer-Nameki, L.~Bhardwaj, and J.~Oh, ``{The Global Form of
  Flavor Symmetries and 2-Group Symmetries in 5d SCFTs},''
  \href{http://dx.doi.org/10.21468/SciPostPhys.13.2.024}{{\em SciPost Phys.}
  {\bfseries 13} no.~2, (2022) 024},
  \href{http://arxiv.org/abs/2105.08724}{{\ttfamily arXiv:2105.08724
  [hep-th]}}.

\bibitem{Kim:2020hhh}
H.-C. Kim, M.~Kim, S.-S. Kim, and K.-H. Lee, ``{Bootstrapping BPS spectra of
  5d/6d field theories},''
  \href{http://dx.doi.org/10.1007/JHEP04(2021)161}{{\em JHEP} {\bfseries 04}
  (2021) 161}, \href{http://arxiv.org/abs/2101.00023}{{\ttfamily
  arXiv:2101.00023 [hep-th]}}.

\bibitem{Hayashi:2021pcj}
H.~Hayashi, H.-C. Kim, and K.~Ohmori, ``{6d/5d exceptional gauge theories from
  web diagrams},'' \href{http://dx.doi.org/10.1007/JHEP07(2021)128}{{\em JHEP}
  {\bfseries 07} (2021) 128}, \href{http://arxiv.org/abs/2103.02799}{{\ttfamily
  arXiv:2103.02799 [hep-th]}}.

\bibitem{Kim:2021gyj}
H.-C. Kim, M.~Kim, and S.-S. Kim, ``{5d/6d Wilson loops from blowups},''
  \href{http://dx.doi.org/10.1007/JHEP08(2021)131}{{\em JHEP} {\bfseries 08}
  (2021) 131}, \href{http://arxiv.org/abs/2106.04731}{{\ttfamily
  arXiv:2106.04731 [hep-th]}}.

\bibitem{Hayashi:2018lyv}
H.~Hayashi, S.-S. Kim, K.~Lee, and F.~Yagi, ``{Dualities and 5-brane webs for
  5d rank 2 SCFTs},'' \href{http://dx.doi.org/10.1007/JHEP12(2018)016}{{\em
  JHEP} {\bfseries 12} (2018) 016},
  \href{http://arxiv.org/abs/1806.10569}{{\ttfamily arXiv:1806.10569
  [hep-th]}}.

\bibitem{Apruzzi:2018nre}
F.~Apruzzi, L.~Lin, and C.~Mayrhofer, ``{Phases of 5d SCFTs from M-/F-theory on
  Non-Flat Fibrations},'' \href{http://dx.doi.org/10.1007/JHEP05(2019)187}{{\em
  JHEP} {\bfseries 05} (2019) 187},
  \href{http://arxiv.org/abs/1811.12400}{{\ttfamily arXiv:1811.12400
  [hep-th]}}.

\bibitem{Apruzzi:2019opn}
F.~Apruzzi, C.~Lawrie, L.~Lin, S.~Sch\"afer-Nameki, and Y.-N. Wang, ``{Fibers
  add Flavor, Part I: Classification of 5d SCFTs, Flavor Symmetries and BPS
  States},'' \href{http://dx.doi.org/10.1007/JHEP11(2019)068}{{\em JHEP}
  {\bfseries 11} (2019) 068}, \href{http://arxiv.org/abs/1907.05404}{{\ttfamily
  arXiv:1907.05404 [hep-th]}}.

\bibitem{Apruzzi:2019kgb}
F.~Apruzzi, S.~Schafer-Nameki, and Y.-N. Wang, ``{5d SCFTs from Decoupling and
  Gluing},'' \href{http://dx.doi.org/10.1007/JHEP08(2020)153}{{\em JHEP}
  {\bfseries 08} (2020) 153}, \href{http://arxiv.org/abs/1912.04264}{{\ttfamily
  arXiv:1912.04264 [hep-th]}}.

\bibitem{Saxena:2020ltf}
V.~Saxena, ``{Rank-two 5d SCFTs from M-theory at isolated toric singularities:
  a systematic study},'' \href{http://dx.doi.org/10.1007/JHEP04(2020)198}{{\em
  JHEP} {\bfseries 04} (2020) 198},
  \href{http://arxiv.org/abs/1911.09574}{{\ttfamily arXiv:1911.09574
  [hep-th]}}.

\bibitem{Eckhard:2020jyr}
J.~Eckhard, S.~Sch\"afer-Nameki, and Y.-N. Wang, ``{Trifectas for T$_{N}$ in
  5d},'' \href{http://dx.doi.org/10.1007/JHEP07(2020)199}{{\em JHEP} {\bfseries
  07} no.~07, (2020) 199}, \href{http://arxiv.org/abs/2004.15007}{{\ttfamily
  arXiv:2004.15007 [hep-th]}}.

\bibitem{Morrison:2020ool}
D.~R. Morrison, S.~Schafer-Nameki, and B.~Willett, ``{Higher-Form Symmetries in
  5d},'' \href{http://dx.doi.org/10.1007/JHEP09(2020)024}{{\em JHEP} {\bfseries
  09} (2020) 024}, \href{http://arxiv.org/abs/2005.12296}{{\ttfamily
  arXiv:2005.12296 [hep-th]}}.

\bibitem{Hubner:2020uvb}
M.~Hubner, ``{5d SCFTs from (E$_{n}$, E$_{m}$) conformal matter},''
  \href{http://dx.doi.org/10.1007/JHEP12(2020)014}{{\em JHEP} {\bfseries 12}
  (2020) 014}, \href{http://arxiv.org/abs/2006.01694}{{\ttfamily
  arXiv:2006.01694 [hep-th]}}.

\bibitem{Albertini:2020mdx}
F.~Albertini, M.~Del~Zotto, I.~n. Garc\'\i{}a~Etxebarria, and S.~S. Hosseini,
  ``{Higher Form Symmetries and M-theory},''
  \href{http://dx.doi.org/10.1007/JHEP12(2020)203}{{\em JHEP} {\bfseries 12}
  (2020) 203}, \href{http://arxiv.org/abs/2005.12831}{{\ttfamily
  arXiv:2005.12831 [hep-th]}}.

\bibitem{Closset:2020scj}
C.~Closset, S.~Schafer-Nameki, and Y.-N. Wang, ``{Coulomb and Higgs Branches
  from Canonical Singularities: Part 0},''
  \href{http://dx.doi.org/10.1007/JHEP02(2021)003}{{\em JHEP} {\bfseries 02}
  (2021) 003}, \href{http://arxiv.org/abs/2007.15600}{{\ttfamily
  arXiv:2007.15600 [hep-th]}}.

\bibitem{Braun:2021lzt}
A.~P. Braun, J.~Chen, B.~Haghighat, M.~Sperling, and S.~Yang, ``{Fibre-base
  duality of 5d KK theories},''
  \href{http://dx.doi.org/10.1007/JHEP05(2021)200}{{\em JHEP} {\bfseries 05}
  (2021) 200}, \href{http://arxiv.org/abs/2103.06066}{{\ttfamily
  arXiv:2103.06066 [hep-th]}}.

\bibitem{Tian:2021cif}
J.~Tian and Y.-N. Wang, ``{5D and 6D SCFTs from $\mathbb{C}^3$ orbifolds},''
  \href{http://dx.doi.org/10.21468/SciPostPhys.12.4.127}{{\em SciPost Phys.}
  {\bfseries 12} no.~4, (2022) 127},
  \href{http://arxiv.org/abs/2110.15129}{{\ttfamily arXiv:2110.15129
  [hep-th]}}.

\bibitem{Closset:2021lwy}
C.~Closset, S.~Sch\"afer-Nameki, and Y.-N. Wang, ``{Coulomb and Higgs branches
  from canonical singularities. Part I. Hypersurfaces with smooth Calabi-Yau
  resolutions},'' \href{http://dx.doi.org/10.1007/JHEP04(2022)061}{{\em JHEP}
  {\bfseries 04} (2022) 061}, \href{http://arxiv.org/abs/2111.13564}{{\ttfamily
  arXiv:2111.13564 [hep-th]}}.

\bibitem{Jefferson:2017ahm}
P.~Jefferson, H.-C. Kim, C.~Vafa, and G.~Zafrir, ``{Towards Classification of
  5d SCFTs: Single Gauge Node},''
  \href{http://arxiv.org/abs/1705.05836}{{\ttfamily arXiv:1705.05836
  [hep-th]}}.

\bibitem{Ashok:2017bld}
S.~K. Ashok, M.~Billo, E.~Dell'Aquila, M.~Frau, V.~Gupta, R.~R. John, and
  A.~Lerda, ``{Surface operators in 5d gauge theories and duality relations},''
  \href{http://dx.doi.org/10.1007/JHEP05(2018)046}{{\em JHEP} {\bfseries 05}
  (2018) 046}, \href{http://arxiv.org/abs/1712.06946}{{\ttfamily
  arXiv:1712.06946 [hep-th]}}.

\bibitem{Cheng:2018wll}
S.~Cheng and S.-S. Kim, ``{Refined topological vertex for a 5D Sp(N) gauge
  theories with antisymmetric matter},''
  \href{http://dx.doi.org/10.1103/PhysRevD.104.086004}{{\em Phys. Rev. D}
  {\bfseries 104} no.~8, (2021) 086004},
  \href{http://arxiv.org/abs/1809.00629}{{\ttfamily arXiv:1809.00629
  [hep-th]}}.

\bibitem{Cabrera:2018jxt}
S.~Cabrera, A.~Hanany, and F.~Yagi, ``{Tropical Geometry and Five Dimensional
  Higgs Branches at Infinite Coupling},''
  \href{http://dx.doi.org/10.1007/JHEP01(2019)068}{{\em JHEP} {\bfseries 01}
  (2019) 068}, \href{http://arxiv.org/abs/1810.01379}{{\ttfamily
  arXiv:1810.01379 [hep-th]}}.

\bibitem{Bastian:2018fba}
B.~Bastian, S.~Hohenegger, A.~Iqbal, and S.-J. Rey, ``{Five-Dimensional Gauge
  Theories from Shifted Web Diagrams},''
  \href{http://dx.doi.org/10.1103/PhysRevD.99.046012}{{\em Phys. Rev. D}
  {\bfseries 99} no.~4, (2019) 046012},
  \href{http://arxiv.org/abs/1810.05109}{{\ttfamily arXiv:1810.05109
  [hep-th]}}.

\bibitem{Chaney:2018gjc}
A.~Chaney and C.~F. Uhlemann, ``{On minimal Type IIB AdS$_{6}$ solutions with
  commuting 7-branes},'' \href{http://dx.doi.org/10.1007/JHEP12(2018)110}{{\em
  JHEP} {\bfseries 12} (2018) 110},
  \href{http://arxiv.org/abs/1810.10592}{{\ttfamily arXiv:1810.10592
  [hep-th]}}.

\bibitem{Closset:2018bjz}
C.~Closset, M.~Del~Zotto, and V.~Saxena, ``{Five-dimensional SCFTs and gauge
  theory phases: an M-theory/type IIA perspective},''
  \href{http://dx.doi.org/10.21468/SciPostPhys.6.5.052}{{\em SciPost Phys.}
  {\bfseries 6} no.~5, (2019) 052},
  \href{http://arxiv.org/abs/1812.10451}{{\ttfamily arXiv:1812.10451
  [hep-th]}}.

\bibitem{Hayashi:2019yxj}
H.~Hayashi, S.-S. Kim, K.~Lee, and F.~Yagi, ``{Rank-3 antisymmetric matter on
  5-brane webs},'' \href{http://dx.doi.org/10.1007/JHEP05(2019)133}{{\em JHEP}
  {\bfseries 05} (2019) 133}, \href{http://arxiv.org/abs/1902.04754}{{\ttfamily
  arXiv:1902.04754 [hep-th]}}.

\bibitem{Apruzzi:2019vpe}
F.~Apruzzi, C.~Lawrie, L.~Lin, S.~Sch\"afer-Nameki, and Y.-N. Wang, ``{5d
  Superconformal Field Theories and Graphs},''
  \href{http://dx.doi.org/10.1016/j.physletb.2019.135077}{{\em Phys. Lett. B}
  {\bfseries 800} (2020) 135077},
  \href{http://arxiv.org/abs/1906.11820}{{\ttfamily arXiv:1906.11820
  [hep-th]}}.

\bibitem{Kim:2019dqn}
H.-C. Kim, S.-S. Kim, and K.~Lee, ``{Higgsing and twisting of 6d D$_{N}$ gauge
  theories},'' \href{http://dx.doi.org/10.1007/JHEP10(2020)014}{{\em JHEP}
  {\bfseries 10} (2020) 014}, \href{http://arxiv.org/abs/1908.04704}{{\ttfamily
  arXiv:1908.04704 [hep-th]}}.

\bibitem{Kim:2019uqw}
J.~Kim, S.-S. Kim, K.-H. Lee, K.~Lee, and J.~Song, ``{Instantons from
  Blow-up},'' \href{http://dx.doi.org/10.1007/JHEP11(2019)092}{{\em JHEP}
  {\bfseries 11} (2019) 092}, \href{http://arxiv.org/abs/1908.11276}{{\ttfamily
  arXiv:1908.11276 [hep-th]}}. [Erratum: JHEP 06, 124 (2020)].

\bibitem{Uhlemann:2019ypp}
C.~F. Uhlemann, ``{Exact results for 5d SCFTs of long quiver type},''
  \href{http://dx.doi.org/10.1007/JHEP11(2019)072}{{\em JHEP} {\bfseries 11}
  (2019) 072}, \href{http://arxiv.org/abs/1909.01369}{{\ttfamily
  arXiv:1909.01369 [hep-th]}}.

\bibitem{Apruzzi:2019enx}
F.~Apruzzi, C.~Lawrie, L.~Lin, S.~Sch\"afer-Nameki, and Y.-N. Wang, ``{Fibers
  add Flavor, Part II: 5d SCFTs, Gauge Theories, and Dualities},''
  \href{http://dx.doi.org/10.1007/JHEP03(2020)052}{{\em JHEP} {\bfseries 03}
  (2020) 052}, \href{http://arxiv.org/abs/1909.09128}{{\ttfamily
  arXiv:1909.09128 [hep-th]}}.

\bibitem{Cota:2019cjx}
C.~F. Cota, A.~Klemm, and T.~Schimannek, ``{Topological strings on genus one
  fibered Calabi-Yau 3-folds and string dualities},''
  \href{http://dx.doi.org/10.1007/JHEP11(2019)170}{{\em JHEP} {\bfseries 11}
  (2019) 170}, \href{http://arxiv.org/abs/1910.01988}{{\ttfamily
  arXiv:1910.01988 [hep-th]}}.

\bibitem{Gu:2019pqj}
J.~Gu, B.~Haghighat, A.~Klemm, K.~Sun, and X.~Wang, ``{Elliptic blowup
  equations for 6d SCFTs. Part III. E-strings, M-strings and chains},''
  \href{http://dx.doi.org/10.1007/JHEP07(2020)135}{{\em JHEP} {\bfseries 07}
  (2020) 135}, \href{http://arxiv.org/abs/1911.11724}{{\ttfamily
  arXiv:1911.11724 [hep-th]}}.

\bibitem{Uhlemann:2019lge}
C.~F. Uhlemann, ``{AdS$_6$/CFT$_5$ with O7-planes},''
  \href{http://dx.doi.org/10.1007/JHEP04(2020)113}{{\em JHEP} {\bfseries 04}
  (2020) 113}, \href{http://arxiv.org/abs/1912.09716}{{\ttfamily
  arXiv:1912.09716 [hep-th]}}.

\bibitem{Hayashi:2019jvx}
H.~Hayashi, S.-S. Kim, K.~Lee, and F.~Yagi, ``{Complete prepotential for 5d $
  \mathcal{N} $ = 1 superconformal field theories},''
  \href{http://dx.doi.org/10.1007/JHEP02(2020)074}{{\em JHEP} {\bfseries 02}
  (2020) 074}, \href{http://arxiv.org/abs/1912.10301}{{\ttfamily
  arXiv:1912.10301 [hep-th]}}.

\bibitem{Closset:2019juk}
C.~Closset and M.~Del~Zotto, ``{On 5d SCFTs and their BPS quivers. Part I:
  B-branes and brane tilings},''
  \href{http://arxiv.org/abs/1912.13502}{{\ttfamily arXiv:1912.13502
  [hep-th]}}.

\bibitem{Bourget:2020gzi}
A.~Bourget, J.~F. Grimminger, A.~Hanany, M.~Sperling, and Z.~Zhong, ``{Magnetic
  Quivers from Brane Webs with O5 Planes},''
  \href{http://dx.doi.org/10.1007/JHEP07(2020)204}{{\em JHEP} {\bfseries 07}
  (2020) 204}, \href{http://arxiv.org/abs/2004.04082}{{\ttfamily
  arXiv:2004.04082 [hep-th]}}.

\bibitem{Crichigno:2020ouj}
P.~M. Crichigno and D.~Jain, ``{The 5d Superconformal Index at Large $N$ and
  Black Holes},'' \href{http://dx.doi.org/10.1007/JHEP09(2020)124}{{\em JHEP}
  {\bfseries 09} (2020) 124}, \href{http://arxiv.org/abs/2005.00550}{{\ttfamily
  arXiv:2005.00550 [hep-th]}}.

\bibitem{Uhlemann:2020bek}
C.~F. Uhlemann, ``{Wilson loops in 5d long quiver gauge theories},''
  \href{http://dx.doi.org/10.1007/JHEP09(2020)145}{{\em JHEP} {\bfseries 09}
  (2020) 145}, \href{http://arxiv.org/abs/2006.01142}{{\ttfamily
  arXiv:2006.01142 [hep-th]}}.

\bibitem{Gu:2020fem}
J.~Gu, B.~Haghighat, A.~Klemm, K.~Sun, and X.~Wang, ``{Elliptic blowup
  equations for 6d SCFTs. Part IV. Matters},''
  \href{http://dx.doi.org/10.1007/JHEP11(2021)090}{{\em JHEP} {\bfseries 11}
  (2021) 090}, \href{http://arxiv.org/abs/2006.03030}{{\ttfamily
  arXiv:2006.03030 [hep-th]}}.

\bibitem{Collinucci:2020jqd}
A.~Collinucci and R.~Valandro, ``{The role of U(1)\textquoteright{}s in 5d
  theories, Higgs branches, and geometry},''
  \href{http://dx.doi.org/10.1007/JHEP10(2020)178}{{\em JHEP} {\bfseries 10}
  (2020) 178}, \href{http://arxiv.org/abs/2006.15464}{{\ttfamily
  arXiv:2006.15464 [hep-th]}}.

\bibitem{Akhond:2020vhc}
M.~Akhond, F.~Carta, S.~Dwivedi, H.~Hayashi, S.-S. Kim, and F.~Yagi,
  ``{Five-brane webs, Higgs branches and unitary/orthosymplectic magnetic
  quivers},'' \href{http://dx.doi.org/10.1007/JHEP12(2020)164}{{\em JHEP}
  {\bfseries 12} (2020) 164}, \href{http://arxiv.org/abs/2008.01027}{{\ttfamily
  arXiv:2008.01027 [hep-th]}}.

\bibitem{vanBeest:2020kou}
M.~van Beest, A.~Bourget, J.~Eckhard, and S.~Schafer-Nameki, ``{(Symplectic)
  Leaves and (5d Higgs) Branches in the Poly(go)nesian Tropical Rain Forest},''
  \href{http://dx.doi.org/10.1007/JHEP11(2020)124}{{\em JHEP} {\bfseries 11}
  (2020) 124}, \href{http://arxiv.org/abs/2008.05577}{{\ttfamily
  arXiv:2008.05577 [hep-th]}}.

\bibitem{Apruzzi:2020zot}
F.~Apruzzi, M.~Dierigl, and L.~Lin, ``{The fate of discrete 1-form symmetries
  in 6d},'' \href{http://dx.doi.org/10.21468/SciPostPhys.12.2.047}{{\em SciPost
  Phys.} {\bfseries 12} no.~2, (2022) 047},
  \href{http://arxiv.org/abs/2008.09117}{{\ttfamily arXiv:2008.09117
  [hep-th]}}.

\bibitem{Fluder:2020pym}
M.~Fluder and C.~F. Uhlemann, ``{Evidence for a 5d F-theorem},''
  \href{http://dx.doi.org/10.1007/JHEP02(2021)192}{{\em JHEP} {\bfseries 02}
  (2021) 192}, \href{http://arxiv.org/abs/2011.00006}{{\ttfamily
  arXiv:2011.00006 [hep-th]}}.

\bibitem{Bergman:2020myx}
O.~Bergman and D.~Rodr\'\i{}guez-G\'omez, ``{The Cat\textquoteright{}s Cradle:
  deforming the higher rank E$_{1}$ and $ {\tilde{E}}_1 $ theories},''
  \href{http://dx.doi.org/10.1007/JHEP02(2021)122}{{\em JHEP} {\bfseries 02}
  (2021) 122}, \href{http://arxiv.org/abs/2011.05125}{{\ttfamily
  arXiv:2011.05125 [hep-th]}}.

\bibitem{VanBeest:2020kxw}
M.~Van~Beest, A.~Bourget, J.~Eckhard, and S.~Sch\"afer-Nameki, ``{(5d RG-flow)
  Trees in the Tropical Rain Forest},''
  \href{http://dx.doi.org/10.1007/JHEP03(2021)241}{{\em JHEP} {\bfseries 03}
  (2021) 241}, \href{http://arxiv.org/abs/2011.07033}{{\ttfamily
  arXiv:2011.07033 [hep-th]}}.

\bibitem{Closset:2020afy}
C.~Closset, S.~Giacomelli, S.~Schafer-Nameki, and Y.-N. Wang, ``{5d and 4d
  SCFTs: Canonical Singularities, Trinions and S-Dualities},''
  \href{http://dx.doi.org/10.1007/JHEP05(2021)274}{{\em JHEP} {\bfseries 05}
  (2021) 274}, \href{http://arxiv.org/abs/2012.12827}{{\ttfamily
  arXiv:2012.12827 [hep-th]}}.

\bibitem{Hayashi:2020hhb}
H.~Hayashi and R.-D. Zhu, ``{More on topological vertex formalism for 5-brane
  webs with O5-plane},'' \href{http://dx.doi.org/10.1007/JHEP04(2021)292}{{\em
  JHEP} {\bfseries 04} (2021) 292},
  \href{http://arxiv.org/abs/2012.13303}{{\ttfamily arXiv:2012.13303
  [hep-th]}}.

\bibitem{Kim:2021cua}
H.-C. Kim, M.~Kim, and S.-S. Kim, ``{Topological vertex for 6d SCFTs with
  $\mathbb{Z}_2$-twist},''
  \href{http://dx.doi.org/10.1007/JHEP03(2021)132}{{\em JHEP} {\bfseries 03}
  (2021) 132}, \href{http://arxiv.org/abs/2101.01030}{{\ttfamily
  arXiv:2101.01030 [hep-th]}}.

\bibitem{Duan:2021ges}
Z.~Duan, K.~Lee, J.~Nahmgoong, and X.~Wang, ``{Twisted 6d (2, 0) SCFTs on a
  circle},'' \href{http://dx.doi.org/10.1007/JHEP07(2021)179}{{\em JHEP}
  {\bfseries 07} (2021) 179}, \href{http://arxiv.org/abs/2103.06044}{{\ttfamily
  arXiv:2103.06044 [hep-th]}}.

\bibitem{Cvetic:2021sxm}
M.~Cvetic, M.~Dierigl, L.~Lin, and H.~Y. Zhang, ``{Higher-form symmetries and
  their anomalies in M-/F-theory duality},''
  \href{http://dx.doi.org/10.1103/PhysRevD.104.126019}{{\em Phys. Rev. D}
  {\bfseries 104} no.~12, (2021) 126019},
  \href{http://arxiv.org/abs/2106.07654}{{\ttfamily arXiv:2106.07654
  [hep-th]}}.

\bibitem{vanBeest:2021xyt}
M.~van Beest and S.~Giacomelli, ``{Connecting 5d Higgs branches via
  Fayet-Iliopoulos deformations},''
  \href{http://dx.doi.org/10.1007/JHEP12(2021)202}{{\em JHEP} {\bfseries 12}
  (2021) 202}, \href{http://arxiv.org/abs/2110.02872}{{\ttfamily
  arXiv:2110.02872 [hep-th]}}.

\bibitem{Acharya:2021jsp}
B.~Acharya, N.~Lambert, M.~Najjar, E.~E. Svanes, and J.~Tian, ``{Gauging
  discrete symmetries of T$_{N}$-theories in five dimensions},''
  \href{http://dx.doi.org/10.1007/JHEP04(2022)114}{{\em JHEP} {\bfseries 04}
  (2022) 114}, \href{http://arxiv.org/abs/2110.14441}{{\ttfamily
  arXiv:2110.14441 [hep-th]}}.

\bibitem{Apruzzi:2021mlh}
F.~Apruzzi, L.~Bhardwaj, D.~S.~W. Gould, and S.~Schafer-Nameki, ``{2-Group
  symmetries and their classification in 6d},''
  \href{http://dx.doi.org/10.21468/SciPostPhys.12.3.098}{{\em SciPost Phys.}
  {\bfseries 12} no.~3, (2022) 098},
  \href{http://arxiv.org/abs/2110.14647}{{\ttfamily arXiv:2110.14647
  [hep-th]}}.

\bibitem{Saidi:2021kan}
E.~H. Saidi and L.~B. Drissi, ``{5D N=1 super QFT: Symplectic quivers},''
  \href{http://dx.doi.org/10.1016/j.nuclphysb.2021.115632}{{\em Nucl. Phys. B}
  {\bfseries 974} (2022) 115632},
  \href{http://arxiv.org/abs/2112.04695}{{\ttfamily arXiv:2112.04695
  [hep-th]}}.

\bibitem{Kim:2021fxx}
H.-C. Kim, S.-S. Kim, and K.~Lee, ``{S-foldings of 5d SCFTs},''
  \href{http://dx.doi.org/10.1007/JHEP05(2022)178}{{\em JHEP} {\bfseries 05}
  (2022) 178}, \href{http://arxiv.org/abs/2112.14550}{{\ttfamily
  arXiv:2112.14550 [hep-th]}}.

\bibitem{DelZotto:2022fnw}
M.~Del~Zotto, J.~J. Heckman, S.~N. Meynet, R.~Moscrop, and H.~Y. Zhang,
  ``{Higher symmetries of 5D orbifold SCFTs},''
  \href{http://dx.doi.org/10.1103/PhysRevD.106.046010}{{\em Phys. Rev. D}
  {\bfseries 106} no.~4, (2022) 046010},
  \href{http://arxiv.org/abs/2201.08372}{{\ttfamily arXiv:2201.08372
  [hep-th]}}.

\bibitem{Kim:2022dbr}
S.-S. Kim and X.-Y. Wei, ``{Refined topological vertex with ON-planes},''
  \href{http://dx.doi.org/10.1007/JHEP08(2022)006}{{\em JHEP} {\bfseries 08}
  (2022) 006}, \href{http://arxiv.org/abs/2201.12264}{{\ttfamily
  arXiv:2201.12264 [hep-th]}}.

\bibitem{Akhond:2022jts}
M.~Akhond, F.~Carta, S.~Dwivedi, H.~Hayashi, S.-S. Kim, and F.~Yagi,
  ``{Exploring the orthosymplectic zoo},''
  \href{http://dx.doi.org/10.1007/JHEP05(2022)054}{{\em JHEP} {\bfseries 05}
  (2022) 054}, \href{http://arxiv.org/abs/2203.01951}{{\ttfamily
  arXiv:2203.01951 [hep-th]}}.

\bibitem{Cvetic:2022imb}
M.~Cveti\v{c}, J.~J. Heckman, M.~H\"ubner, and E.~Torres, ``{0-Form, 1-Form and
  2-Group Symmetries via Cutting and Gluing of Orbifolds},''
  \href{http://arxiv.org/abs/2203.10102}{{\ttfamily arXiv:2203.10102
  [hep-th]}}.

\bibitem{DeMarco:2022dgh}
M.~De~Marco, A.~Sangiovanni, and R.~Valandro, ``{5d Higgs Branches from
  M-theory on quasi-homogeneous cDV threefold singularities},''
  \href{http://arxiv.org/abs/2205.01125}{{\ttfamily arXiv:2205.01125
  [hep-th]}}.

\bibitem{Jia:2022dra}
Q.~Jia and P.~Yi, ``{Holonomy Saddles and 5d BPS Quivers},''
  \href{http://arxiv.org/abs/2208.14579}{{\ttfamily arXiv:2208.14579
  [hep-th]}}.

\bibitem{Witten:1997kz}
E.~Witten, ``{New 'gauge' theories in six-dimensions},''
  \href{http://dx.doi.org/10.1088/1126-6708/1998/01/001}{{\em JHEP} {\bfseries
  01} (1998) 001}, \href{http://arxiv.org/abs/hep-th/9710065}{{\ttfamily
  arXiv:hep-th/9710065}}.

\end{thebibliography}\endgroup

\end{document}